\documentclass[smallextended]{svjour3}

\usepackage[pdftex,pagebackref=true]{hyperref}

\date{Draft, 2010}

\usepackage{natbib}

\usepackage{mathptmx}

\usepackage{amssymb}

\usepackage{multirow}

\usepackage{framed}

\usepackage{graphicx}

\usepackage{algorithm}
\floatname{algorithm}{Alg.}
\usepackage{algorithmic}

\title{Generate Descriptive Social Networks for Large Populations from Available Observations}
\subtitle{A Novel Methodology and a Generator}
\author{Samuel Thiriot}

\institute{Samuel Thiriot \at IRIT-UT, 
	\\Universit\'e des Sciences Sociales, Toulouse, France
	\\ \email{samuel.thiriot@res-ear.ch}
}

\begin{document}

\bibliographystyle{spbasic} 

\maketitle

% keywords: social networks; random network generator; agent-based modelling; social phenomena; diffusion of information; epidemics

% should be about 100-150 words
\begin{abstract}
When modeling a social dynamics with an agent-oriented approach, researchers have to describe the structure of interactions within the population. Given the intractability of extensive network collecting, they rely on random network generators that are supposed to explore the space of plausible networks. We first identify the needs of modelers, including placing heterogeneous agents on the network given their attributes and differentiating the various types of social links that lead to different interactions. We point out the existence of data in the form of scattered statistics and qualitative observations, that should be used to parameter the generator. \newline
We propose a new approach peculiar to agent-based modeling, in which we will generate social links from individuals' observed attributes, and return them as a multiplex network. Interdependencies between socioeconomic attributes, and generative rules, are encoded as Bayesian networks. A methodology guides modelers through the formalization of these parameters. This approach is illustrated by describing the structure of interactions that supports diffusion of contraceptive solutions in rural Kenya.
\end{abstract}


\section*{Outline}

In section~\ref{indoc:section1}, we clarify the research question of random network generation for agent-based modeling. We highlight several needs of modelers, review the existing approaches, and justify the need of a new approach peculiar to agent-based modeling. We propose another approach (part~\ref{indoc:section2}) that enables modelers to take into account their limited knowledge on the structure of interaction to generate networks, by reusing robust evidence on social selection processes. As we don't pretend to generate social networks in general, but to describe a real network from field observations, we propose a methodology that helps modelers to formalized observations in an uniform representation (\ref{indoc:user_methodo}). This methodology is illustrated with an application to the structure of interactions about contraceptive solutions in Kenya.

In part~\ref{indoc:generation}, we detail the principle and implementation of the generative algorithm. Its ability to describe a complex social structure such as the family structure in Kenya by taking into account spatial location and affiliations is demonstrated. By analyzing the networks generated for the Kenya case (\ref{indoc:experiments}), we discuss several properties of this network generator. For instance, a minimal population size is required for the population to be statistically representative of the real one.

\section{Needs of agent-based modeling and limits of existing approaches\label{indoc:section1}} 

\subsection{Social networks for describing the structure of interactions}

\subsubsection{Importance of the network of interactions}

When a modeler creates an agent-based model of a given social phenomenon, s/he faces inevitably the problem of the structure of interactions, that can be summarized as: ``who interacts with who within the population~?''. It is nowadays common to represent these interactions as a so-called ``social network'' $X(A,L)$. Individuals (or their abstraction, the agents) are represented as vertices~$A$ and interactions as links~$L$. While this network is operationally used to define the \textit{structure of interactions}, it is often considered to be similar to to the network of relationships, thus being called ``social network'' in a generic way. 

The interplay between the social network and individuals' state and behavior is bidirectional. Most of the time, modelers focus on the \textit{influence of the social network on individuals' state and behavior}. The definition of the network of interactions $X$ comes to choose of particular network in the large set of all possible networks $\mathcal{X}$, that will be used to support interactions during a simulation run. The typical approach, as observed in specialized journals or international conferences, consists in using one of the most famous network generators to create this structure. The Watts-Strogatz small-worlds, and Barabasi-Albert scale-free networks, are by far the most used network generators \citep[pp~219-257]{samuel_thiriot:bib_sma_simulation:phan_2007_1}. Simulations are thus driven on networks generated by each of these two generators, and their results compared. The reverse process, that is studying the \textit{impact of individuals' behavior on the structure}, constitutes a very different problematic of research, in which the object of the study is not the social dynamics but the dynamics of the network itself. Modelers dealing with this problematic have to develop their own model of network evolution. Anyway, they still need a zero-state network for starting the process (see for instance ). In both cases, modelers have to describe the network of interactions in their population.

% Typically, dynamics are observed to depend on the network generator. 

It was shown for a lot of models that this network has a dramatic influence on the collective dynamics. That is the case for the diffusion of knowledge \citep{samuel_thiriot:bib_sma_simulation:cowan_2004_1,samuel_thiriot:bib_sma_simulation:valente_1995_1,samuel_thiriot:bib_sma_simulation:wu_2004_1}, the diffusion of innovations \citep{samuel_thiriot:bib_sma_simulation:abrahamson_1997_1,samuel_thiriot:bib_perso:thiriot_2008_2} and various other processes. That sensitivity makes the structure of interactions a parameter of very first importance, not only for one given agent-based model, but for \textit{most} individual-centric models \citep{samuel_thiriot:bib_sma_simulation:carley_2009_1}. Given evidence on the sensitivity of simulation to this network, the plausibility of the network directly impacts the plausibility of the simulation. The creation of descriptive networks is thus a fundamental need for modelers.

% Given their importance, the networks defining the structure of interactions may and should be studied as a modelling problematics.

\subsubsection{Illustration of the problematics: diffusion of contraception in Kenya}

We propose to illustrate the problematic of modelers with a field case: the modeling of the structure of interactions that influence the diffusion of contraceptive solutions in rural Kenya. One benefit of this case is its singularity: polygyny is allowed in this Luo culture, the demographical transition is still at its beginning and long-distance travels are quite rare. Let us precise that \textit{we collected no data from our own}; observations on this social phenomenon are all retrieved from published studies and public statistics. The Kenya illustration will be used during the whole all paper. In order to improve lisibility, parts related to this specific case will be framed.

\begin{framed}
Diffusion of a new practice like contraceptive use heavily depends on the structure of interactions in the population \citep{samuel_thiriot:bib_customer_value:rogers_2003_1}. During the diffusion of contraceptive pills in rural Kenya (Nyanza), field observations indicate that this structure has two main effects that change the probability to adopt \citep{samuel_thiriot:bib_customer_value:behrman_2002_1,samuel_thiriot:bib_customer_value:kohler_2001_1}. \textit{Social interactions support information transmission} on this topic. Women daily discuss the supposed dangers of pills, thus reinforcing the fears that adoption may lead to malformed children \citep{samuel_thiriot:bib_customer_value:rutenberg_1997_1}. However, this topic remains sensitive and private. They avoid to discuss the topic with their own mother or mother-in-law, who are often traditionalist and against this practice. Women prefer to discuss the topic only with people they trust, that is their sisters, their long-term friends or some colleagues. We know that social networks in sub-Saharan Africa are quite gendered: women talk with women, men with men \citep{samuel_thiriot:bib_customer_value:watkins_2004_1}.  The social structure also leads to \textit{normative influence}. Women hesitate to adopt contraceptive solutions, because they don't like to act against their mother's advice. In the same way, young men are reticent to allow their wives to adopt contraceptive pills: their fathers and mother disagree with their practice, and their social perception is mainly based on having many children. Normative influence may be positive as well, when friends of near family have adopted the solution. Final adoption of a contraception solution finally requires a negotiation between the woman and her husband. 

The characteristics of individuals also impact their behavior. Without surprise, women in age to procreate participate in more discussions that others \citep{samuel_thiriot:bib_customer_value:watkins_1995_2}. Other factors influence the adoption process. For instance, people going at church are less favorable to contraception. 
\end{framed}

In such a situation, modelers would like to describe a plausible structure of interactions, as evidence from both simulation and from the field proved strong sensitivity to this network). They also would like to use qualitative observations from the field for constraining the network that will be used in their simulation. Surprisingly, no network generator enables them to take that kind of field observation into account.

\subsubsection{Purpose of network generators}

% on ne peut pas collecter

The real social phenomenon occurs on a real structure of interactions $X^{real}$. Modelers cannot collect extensively this network for various reasons: data collecting requires costly individual interviews, answers from individuals are always biased due to reminding problems \citep{samuel_thiriot:bib_psycho:brewer_2000_1}. As modelers wish to study diffusion at the regional or national scale, such a data collecting is actually intractable. Sampling this networks also would lead to strong biases in the results \citep{samuel_thiriot:bib_psycho:scott_2001_1,samuel_thiriot:bib_psycho:alba_1982_1,samuel_thiriot:bib_psycho:frank_1978_1}. In short, \textit{modelers don't want nor can collect the real network of interactions from the field}.

In such a situation, modelers typically use random network generators to create the network of interaction $X$. 
Network generators are generative algorithms that are supposed to restrict the space of possible networks $\mathcal{X}$ to a subset of plausible networks $\mathcal{X}^{plausible}$, given hypothesis of which networks are ``plausible'' given observations or theories. For instance, the Watts-Strogatz generative algorithm is built on the principle that real networks are characterized by both an high clustering rate and a short average path length, and generates networks that are compliant with these observations. Modelers prefer the use of \textit{random} network generators, that concretely explore the space $\mathcal{X}^{plausible}$ by randomly generating networks compliant with observations. Randomness permits to take into account our lack of knowledge on the real structure of interactions $X^{real}$: as this structure is unknown, randomness enables us to explore the consequences of the possible structures on the simulated dynamics.

The use of network generators also fullfits various implicit needs of modelers. They constitute \textit{tools ready to use out-of-the-box}, that satisfy all the modelers (especially the more and more numerous newcomers  in computational modelling \citep{samuel_thiriot:bib_sma_simulation:harrison_2007_1}) that don't aim to devote much time to the network part of their model. Generators enable the generation of large populations, thus satisfying modelers that are often interested to large scale phenomena (like adoption of novel practices or pandemics). They also enhance the \textit{communicability} of models. When publishing the results of an agent-based model, describing the behavior of agents, their interactions and simulation results already require so much space that detailing an original structure of interactions becomes impossible. Famous generators preserve communicability and replicability of results, as they can be simply cited with one reference in a journal.

These needs explain the use of random network generators that exist today. We add to these needs several properties that an ideal network generator should satisfy. 

% 
% Given the proven sensitivity of collective dynamics to the structure of interactions, random generators should generate plausible networks of interactions. As a consequence, they could and \textit{should} be analyzed as a modelling problematics. 

\subsection{Characteristics of an ideal network generator \label{indoc:peculiar_needs_abm}}

\subsubsection{Flexibility}

% Given evidence of the sensitivity of simulations to the network of interactions, networks generators may and \textit{should} be analyzed as a modelling problematic. As any model, they should be based on \textit{plausible theories} built from \textit{relevant and representative datasets}. 

The concept named ``social networks'' covers many different objects, including relationships networks, interaction networks, support networks and so on. In the original stream of social networks analysis (SNA) \citep{samuel_thiriot:bib_psycho:wasserman_1994_1}, no generic definition of social networks was provided. In practice, the meaning of links in a social network directly depends on the protocol of data collecting. Properties of these networks may be expected to vary as well. Support networks, defined as the individuals from whom one would see advice averages about a degree of 5 \citep{samuel_thiriot:bib_psycho:milardo_1992_1}. When the definition of ``social networks'' is broader, depending to the precise definition, degree may vary from about 250 acquaintances \citep{samuel_thiriot:bib_psycho:killworth_1984_1} up to 5000 \citep{samuel_thiriot:bib_psycho:killworth_1990_1,samuel_thiriot:bib_psycho:desolapool_1978_1}. If such a basic property of networks like the average degree may vary so much, then we can arguably claim that \textit{no generic thing such as ``social networks'' exists}. Social networks are only a metaphor \citep{samuel_thiriot:bib_psycho:breiger_2004_1} that proposes to represent the complex relationships between human beings as links between nodes. 

In agent-based modelling, we are basically interested in \textit{interactions networks}, in which links represent a possibility of interaction between two individuals. Many social phenomena like recommendation processes, opinion dynamics or epidemics occur over stable relationships and could use a relationship network to describe interactions. Word of mouth will rather be interested in the interaction network that describes discussions in the workplace and at school. A model of fads could use the network of people simply viewing others when walking in the street, because vision enables the transmission of adoption or reject of a fad. As a consequence, \textit{a network generator should accept parameters that enable modelers to apply it to the specific network s/he is interested in}.

\subsubsection{Different kinds of links}

Different types of social links don't lead to the same nature not frequency of interaction. For instance, in the case of Kenya, observations prove that all the social links don't lead to the same social influence. From a qualitative viewpoint, some relationships have a stronger normative influence than other. From the quantitative viewpoint, interactions are also more or less frequent given the nature of relationships. These differences are robust accross cultures and social phenomena. The seminal work of Granovetter on the weak and strong ties \citep{samuel_thiriot:bib_psycho:granovetter_1973_1,samuel_thiriot:bib_psycho:granovetter_1983_1} shown that infrequent interactions may lead to stronger changes than frequent ones. Word-of-mouth on products was shown more frequent between colleagues and friends, but more influential between friends even at long distance \citep{samuel_thiriot:bib_customer_value:carl_2006_1}. Studies on Facebook (a website dedicated to social networking) suggest that few links are actually leading to interactions \citep{samuel_thiriot:bib_psycho:golder_2006_1,samuel_thiriot:bib_psycho:lewis_2008_1}.

In order to describe the structure of interactions and its effect on individuals' state and behavior, \textit{an ideal generator should so generate a multiplex network containing the various types of relationships that may lead to different interactions during simulation.}

\subsubsection{Place agents on networks given their characteristics}

Not only the modelers need to describe the structure of interaction, but they also have to position an heterogeneous population of agents in this network. Previous studies prove that the position of agents in a network influence their social influence, and more generally the collective dynamics in its whole (e.g. \citep{samuel_thiriot:bib_sma_simulation:kempe_2003_1}. Evidence from the field indicates that people aren't positioned randomly in the structure of interactions. In seems that people sharing common socioeconomic characteristics are more likely to bond (homophily, see~\ref{indoc:social_selection_processes}). As example, studies on the diffusion of innovations indicate that people having the same attitude towards novelty are often in interaction \citep{samuel_thiriot:bib_customer_value:rogers_2003_1}. To comply with this evidence, \textit{a network generator should position agents in the network given their characteristics.}

\subsubsection{Use qualitative observations and scattered statistics}

The intractability of data collecting is perceived as the main obstacle against the description of large-scale networks. However, many observations are available on this structure of interactions without extensive collecting of the structure itself. In the example of Kenya, we saw that qualitative observations are available on the \textit{structure of interactions} (women mainly discuss with other women and men with men), on \textit{the opportunities for people to meet} and discover each other (people mainly discuss at the workplace, when working at market or working in fields) and on the \textit{nature of interactions supported by each type of relationship} (family links lead to normative influence, while colleagues and friends frequently discuss non-private topics). These qualitative observations complete the \textit{scattered statistics} published by governmental and other institutions, that include statistics on demographics, affiliations of people (attending school, working and in which domain, doing sport, etc.). 

These observations are not complete nor unbiased. However, \textit{they constitute the only available evidence on the idiosyncratic characteristics of the network}. Given the absence of other observations on the structure that the generator is supposed to reproduce, these \textit{qualitative observations and scattered statistics should be used as a parameter for constraining the generated network}.

\subsection{Existing approaches and their limits}

Now that we have defined the needs of Agent-Based Modelling for social networks 
% (summarized in table \ref{tab:requirements_generators})
, we propose an overview of the tremendous activity of social network modelling in various research streams.

\subsubsection{Large Network Analysis}

The most used generators for agent-based models are the Watts-Strogatz and the Barabasi-Albert algorithms \citep{samuel_thiriot:bib_sma_simulation:phan_2007_1}. The first explains the small-world phenomenon, that is the surprising coexistence of a high clustering rate () and a short average path length, by proving that the random creation of links accross a highly clustered network reduces the geodesic distance while keeping clustering high. In a similar way, Barabasi and Albert proposed an explanation of the power-law distribution of degrees observed in many networks \citep{samuel_thiriot:bib_sma_simulation:barabasi_1999_1}, by demonstrating by simulation that a growing network in which new nodes attach preferally to already well-connected nodes reproduce this effect. WS and BA generators are representative of the young field named ``large networks analysis'' or ``the new science of networks'' \citep{samuel_thiriot:bib_sma_simulation:watts_2004_1}. The methodology of this field, mainly driven by physistics and mathematicians, comes to identifying statistical properties shared by many networks then explaining why these properties are so common. This field proposed a large number of generative algorithms to test hypothesis (for an overview of this approach, see \citep{samuel_thiriot:bib_sma_simulation:strogatz_2001_1}, \citep{samuel_thiriot:bib_sma_simulation:albert_2002_1} or \citep{samuel_thiriot:bib_sma_simulation:dorogovtsev_2002_1}).

Trying to apply these models for Kenya reveals their inadequacy. If we decide to use the WS generator to generate the network, we will obtain a network that don't distinguishes the different kinds of interaction. The use of the Barabasi-Albert algorithm would create a population in which most people would have less than 2 acquaintances and some hundreds of neighboors. Such a network seems implausible given sociological observations: everyone, independently to his culture, seems to possess a support network of tens of people, and theories indicate that humans cannot maintain relationships with more than hundreds of persons \citep{samuel_thiriot:bib_psycho:hill_2002_1,samuel_thiriot:bib_psycho:roberts_2009_1}. More generally, models in LNA don't describe the nodes of attributes, nor different kind of links. They also don't enable the use of available observation to parameter the generator. In fact, \textit{LNA just don't tackles the same problematics that ours}: they seek to detect and explain the statistical properties shared by many kinds of networks, while we look for plausible models of social interactions that comply with our observations. 
% If the use of these methods may be understood as a default solution. 

\subsubsection{Social Network Analysis}

In the frame of Social Network Analysis (SNA) \citep{samuel_thiriot:bib_psycho:wasserman_1994_1}, a lot of models were proposed (see \citep{samuel_thiriot:bib_psycho:robins_2001_1} for a synthetic picture). SNA classically starts from an extensive data collecting of a small network, then quantifies the properties of the network and build a model for testing a theory explaining the existence of links. Models are mostly based on small samples of networks, by taking individuals' attributes into account. One of the most famous models proposed by SNA is the p\textasteriskcentered one \citep{samuel_thiriot:bib_sma_simulation:robins_2007_3}, that was shown to be enable the reconstruction of small social networks, like friendship in a classroom given individuals' attributes. This field observed and theorized the processes leading to the existence of link like homophily or transitivity (cf~\ref{indoc:social_selection_processes}).

SNA roots models in real data. However, as their methodology starts from real data, their models are built for being parametrized from observations from the real network. Such models cannot be applied for agent-based modelling. For instance, if we wish to use the p\textasteriskcentered  model for generating a population, we have to determine many parameters like the number of triangles or the reciprocity of links that aren't available in our case, nor than relevant at a large scale (see \citep{samuel_thiriot:bib_sma_simulation:goodreau_2007_1} for an example of application). These models also don't tackle the problem of multiple kinds of relationships. Nevertheless, they proposed insights on the reasons for a link to exist that will be reused in this paper.

\subsubsection{Other approaches\label{indoc:approaches_abm}}

% models like social circles...
Many researchers proposed \textit{individual-centric models of network creation}. Given local rules that govern link creation, these models enable the study of the structure that emerges and (when relevant) its dynamics. As example \citep{samuel_thiriot:bib_sma_simulation:cowan_2004_1} investigated the joint evolution of network and knowledge. Most of these networks are analogical \citep{samuel_thiriot:bib_sma_simulation:edmonds_2005_2} rather than descriptive. The rare models parametered and validate against real networks (e.g.~\citep{samuel_thiriot:bib_sma_simulation:pearson_2006_1}, see \citep{samuel_thiriot:bib_sma_simulation:snijders_2009_1} for an introduction) use a SNA methodology, thus focusing on small networks. No one of these models satisfies the needs listed before. 

\textit{Epidemiology} led to the development of descriptive models of interactions in large populations. For instance, the model detailed in \citep{samuel_thiriot:bib_sma_simulation:stroud_2007_1} takes into account the characteristics of individuals to describe their daily contacts. However, these models are developed for a given population and cannot be extended to another one. The second drawback of these models (given our needs) is that they only focus on the network of interactions susceptible to lead to contagion, while we prefer to give users the freedom to study relationships networks, support networks, family structure and so on.

% handmade
An original solution was recently propose in ABM. As we are looking for realistic networks, why not simply using one of the few \textit{real networks} that were collected from the field~? \citep{samuel_thiriot:bib_sma_simulation:cointet_2007_1} This interesting suggestion fails to match our needs for two main reasons. First, it implies to use one given network to describe another structure of interactions, assuming that these networks are similar. In other words, it suggests we should use a network of phone calls collected in great Brittany (for instance) to describe the structure of interaction on contraceptive pills in Kenya. Secondly, this use wouldn't enable modelers to explore the space of plausible networks, as done with a random generators.

\subsection{The need of a new approach}
% 
% \begin{table}
% \begin{tabular*}{\textwidth}{|p{0.15\textwidth}|p{0.20\textwidth} p{0.20\textwidth} p{0.20\textwidth}|}
% \hline
% 		
% 	& LNA	
% 	& SNA 	
% 	& ABM needs	\\ 
% \hline 
% object modelled	
% 	& any network
% 	& determined by data collecting	
% 	& determined given field observations and the social phenomenon 
% 	\\
% data selected	
% 	& any available network 	
% 	& small samples	of social networks	
% 	& field observations 
% 	\\
% network size	
% 	& large	
% 	& small 
% 	& large	
% 	\\
% model purpose 
% 	& explicative 
% 	& explicative or descriptive 
% 	& descriptive 
% 	\\
% parameters
% 	& statistics on graph properties
% 	& 
% theories on generative process 
% 	& build on purpose \\
% theories 	& general & statistics at the graph scale & \\
% 
% \hline
% \end{tabular*} 
% \caption{Characteristics of research strems on social networks, and the needs of Agent-based Modelling.}
% \end{table} 

The overview of the main research streams highlights that no model satisfies the needs of agent-based modelers. This lack is simply due to the peculiar needs of agent-based modelling: we are willing to describe the structure of interactions in large populations from available observations, while \textit{other fields tackle different problematics}. LNA detects statistical properties on various networks and propose explicative models to explain them. SNA develops models intented to fit collected data on small social networks. Other models explore the consequences of theories on the local processes of network creation for the global generated network. 

% In fact, the need of more descriptive network TODO citer Many authors previously pointed out the need of more descriptive models \citep{samuel_thiriot:bib_sma_simulation:roth_2007_1}.

% parameters
% We need a network generator that may be parameterized with the limited available information. Models from SNA are parameterized from observations on the structure of interactions, that has to be already collected. Models from LNA require statistical properties of networks as parameters, while we don't possess that kind information when modelling a social phenomenon. On the contrary, we have access to information in the form of qualitative observations and scattered statistics that cannot be used with existing solutions.

We claim that \textit{agent-based modelling needs a specific approach to satisfy its peculiar needs}. We already identified the needs that seem us crucial for modelers. An ideal generator should be easily usable as a tool by modelers. It should enable users to use available observations and statistics to constraint the generated networks. Rather than trying to generate a kind of ``social networks'' with weak descriptive power, the generator should be parametrizable for being adapted to the needs of the modeler. The generator should create multiplex networks, with the different kinds of relationships that may lead to different interactions. It should also place agents on the network given their characteristics. 

% This evidence could arguably be analyzed as a fundamental limitation of agent-based simulation. Indeed, we cannot infer anything from simulation results that change dramatically given a parameter that cannot be collected from the field. Even mathematical analysis of individual-centric processes on networks generated by a given algorithm, like the calculus of epidemical threshold for small-world networks [TODO], are of weak interest if these networks aren't descriptive of real ones. However, this limitation seems somewhat to be implicitly accepted in agent-based modelling. In this paper, we attempt to tackle this limitation rather than accepting it by coming back to the fundamental problematics. 

\section{A novel approach\label{indoc:section2}}

\begin{table}[th]
\begin{tabular*}{\textwidth}{|p{0.25\textwidth}|p{0.7\textwidth}|}
\hline
\textbf{need}	&	\textbf{answer}\\
\hline
plausibility 			& the generator is build from observations from the structure of interactions \\
large scale networks		& the generator main be parametered for generating any population size \\
no extensive data collecting	& the approach relies on observations on the probability for agents to link, that are already available and may be collected at low cost \\
randomness	& the generator will use a random component during the generation \\
usable as a tool		& the generator limits the modeler to the introduction of parameters, without any programming required \\
communicability & files used to parameter the generator may be communicated along the source code \\
\hline
flexibility	& \multirow{2}{*}{user choose the kind of links he wants to create} \\
different links &  \\
place agents	& the generator creates the links given agents attributes, thus placing the agents on the network \\
use available information & all possible information may be encoded into parameters \\
\hline
\end{tabular*}
\caption{Our answers to the needs identified for agent-based modelling}
\label{tab:needs_and_answers}
\end{table} 

We propose an approach dedicated to the random generation of networks for computational sociology. This model is build as a tool for models, in order to enable them to generate easily plausible networks from available observations, and given their specific needs. Our answers to the peculiar needs of agent-based modelling are summarized in table~\ref{tab:needs_and_answers}

\subsection{Overview of generator usage} 

\begin{table}[h]
\includegraphics[width=\textwidth]{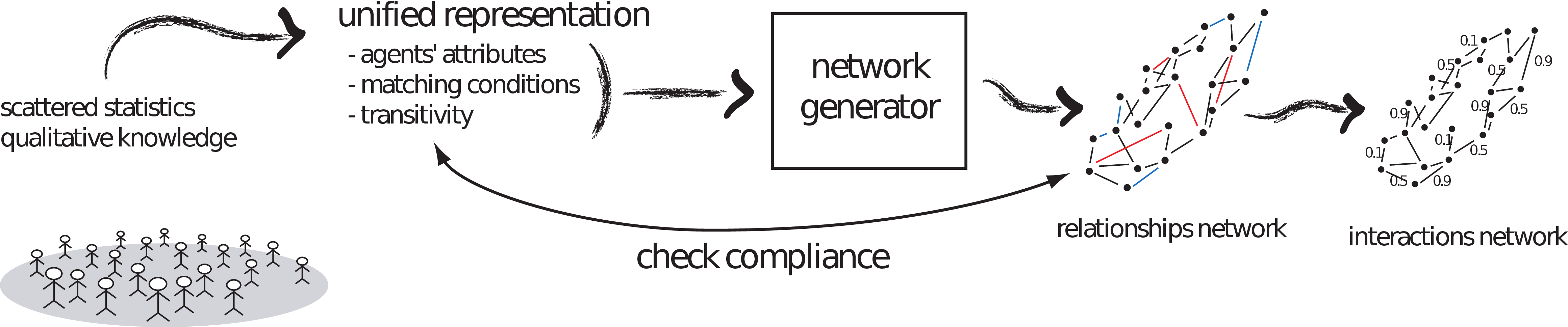}
\caption{Overview of the generating process, from formalization of scattered statistics and qualitative observations using to the actual use of generated networks}
\label{fig:methodo}
\end{table} 

% The purpose of the network generator is to avoid users 

Figure~\ref{fig:methodo} depicts the use of this generator, from formalization of parameters to the actual use of networks. 
\begin{enumerate}
\item \textbf{start from available information:} as said previously, scattered statistics and qualitative observations are available for much social phenomena. We describe more precisely the nature of these observations in \ref{indoc:available_data}.
\item \textbf{formalization of observations:} for being usable by a generator, these observations have to be brought together in a formal representation. When doing so, users also provide as a parameters the types of links and agents' attributes they want in the network. We propose a methodology that helps users to do so. This methodology will be described later in section~\ref{indoc:user_methodo}. 
\item \textbf{Actual generation:} from the user viewpoint, the network generator is only a tool (a black box) that creates networks compliant with the parameters passed as an input. This algorithm and its implementation are detailed in part~\ref{indoc:generation}, as well as the error measure that checks whether the generated network complies with the parameters.
\item \textbf{Generated network:} the network generated by the algorithm is a multiplex network. Each agent (or node) in the network is associated with its characteristics, and positioned on the network depending on them.
\item \textbf{Interaction network:} the user may either directly use the multiplex network for driving his/er simulations, or traduce this multiplex network as a probabilist interaction network.
\end{enumerate}

\subsection{Main principles\label{indoc:principles}}

The basic purpose of a network generator is to constrain the characteristics of the networks it generates from field observations. These observations may either be encoded into the generative algorithm itself or be accepted as a parameter. For instance, the BA algorithm encodes preferential attachment in its process, and accepts the tendency of nodes to attach preferentially as a parameter. Given the needs of modelers enumerated before, we would like to enable modelers to use available information as a parameter. The theory used into the generator should give users the liberty to generate the kind of networks they want, while remaining generic enough to be a tool usable in a quiet easy way. 

\paragraph*{}
We propose the following principle:
\begin{itemize}
\item The generative algorithm itself will be rooted in findings on social selection processes. The generator is based on \textit{two main generative principles}: \textit{creation of links between two individuals given their characteristics} and \textit{creation of links by transitivity}. In~\ref{indoc:social_selection_processes}, we summarize the findings on social selection processes, and explain our interpretation of these findings in order to create a generator. The use of these principles will be later shown generic enough to reproduce a social structure as complex as the structure of family accross a large population.
\item Parameters are exprimed as generative rules that use one of the two generative principles enumerated before. In order to formalize statistics coming from various sources and qualitative observations, we propose to bring these information together using Bayesian Networks. A user methodology guides modelers in this formalization. Illustration to Kenya will demonstrate the flexibility offered by these graphical models. As will be shown for the algorithm itself, Bayesian networks will also facilitate the actual generation of the network.
\end{itemize}

\subsubsection{Observations on social selection processes\label{indoc:social_selection_processes}}

We propose here a synthetic overview of knowledge on social selection processes. We focus on consensual theories that can (and will) be used for building our generator.

\textit{Homophily} refers the tendency of individuals to create relationships with people sharing similar characteristics \citep{samuel_thiriot:bib_psycho:lazarsfeld_1954_1}. This observations was shown to be robust in social networks \citep{samuel_thiriot:bib_psycho:mcpherson_2001_1}. For instance, friends are often of the same gender, similar age and ethnicity. Spouses often belong to the same socioeconomic class. Homophily is often viewed as a preference during social selection processes: people bond because they share similar characteristics. This classical viewpoint on homophily is named \textit{inbreeding homophily}. For instance, friends often have the same age because they studied in the same school or lived in the same district, giving them the possibility to meed and bond. The notion of \textit{assortativity} proposed in LNA subsumes the homophily principle \citep{samuel_thiriot:bib_sma_simulation:newman_2002_2}. Assortativity is defined as tendency of nodes to attach to others that are similar \textit{or different} (dissassortativity) in some way. If assortativity is often measured on the degree of nodes, it may also refer to other characteristics of individuals. Sharing affiliations (event, project or workplace) also increases the probability for two agents to be tied together \citep{samuel_thiriot:bib_psycho:wasserman_1994_1}. 

% reasons we have to meet
However, homophily may also be saw as the consequence of the limited pool of ties available for each individual \citep{samuel_thiriot:bib_psycho:feld_1981_1,samuel_thiriot:bib_psycho:mcpherson_2001_1}. \textit{Baseline homophily} refers to the fact that we do not choose our acquaintances in the \textit{entire} population, but rather in the limited set of people we have the opportunity to meet \citep{samuel_thiriot:bib_psycho:festinger_1950_2,samuel_thiriot:bib_psycho:caplow_1950_1} (see \citep{samuel_thiriot:bib_psycho:mollenhorst_2008_1} for an example of study). Spatial location, or neighborhood, constitutes a first condition for meeting. Despite of technological advances that enable long distance communication, it seems that social support and friendship remain essentially local \citep{samuel_thiriot:bib_psycho:wellman_1990_1} and that these links rarely appear because of electronic solutions \citep{samuel_thiriot:bib_psycho:mayer_2008_1}. 

\textit{Transitivity} refers to the tendency ``friends of friends are also my friends''. Transitivity is both a process and as a statistical indicator. The transitivity process traduces the fact that you have more chances to meet friends of your friends than other people (because you meet them at a party, at you're friend's home, etc). Moreover, you probably share common interests and characteristics with your friends that are also shared by friends of your friends. Thus you have not only more chances to meet friends of friends, but also more chances to appreciate them and bond. The \textit{transitivity statistical indicator} evaluates the number of nodes (N1, N2) in a network that are linked together and with a third node N3. This quantification may be underlied by various processes that include homophily or sharing a common affiliation \citep{samuel_thiriot:bib_psycho:goodreau_2009_1}.  

% To close this overview, we would like to cite two last concepts: social groups and affiliations. Cooley identified several basic groups; the primary group TODO. \cite{samuel_thiriot:bib_psycho:cooley_1909_1}

Contrarily to the graph statistical properties used in LNA, these social selection processes were observed from hundreds of real social networks. In our approach, \textit{we will rely on these observations to build the generative processes encoded in the generator}. We retain two main principles:
\begin{itemize}
\item The probability for agents to be linked together often depends on their attributes. This general rule covers inbreed and baseline homophily as well as affiliations. We will later refer to links created given individuals' attributes as homophily in a larger acceptation: probability for agents to be linked together depends on the similarity, dissimilarity or any relation between their attributes. For instance, the social link ``spouses'' is more probable between agents of different gender. 
\item Transitivity means that a link exists more probably between agents that share a third acquaintance. Transitivity generative rules will enable the description of the fact that people have more chances to meet when related to the same person. More generally, they will also allow the construction of networks in which the existence of links depend on the existence of other links - for instance, the probability for $F$ to be the father of a child $C$ in by far more probable if F is married with $M$, and $M$ is the mother of $C$.
\end{itemize}
 
These two main generative principle only constitute a framework. We will see that field observations may be used to parameter these generative rules.

\subsubsection{Available data\label{indoc:available_data}}

\begin{figure}[tph]\sidecaption
\includegraphics[width=0.7\textwidth]{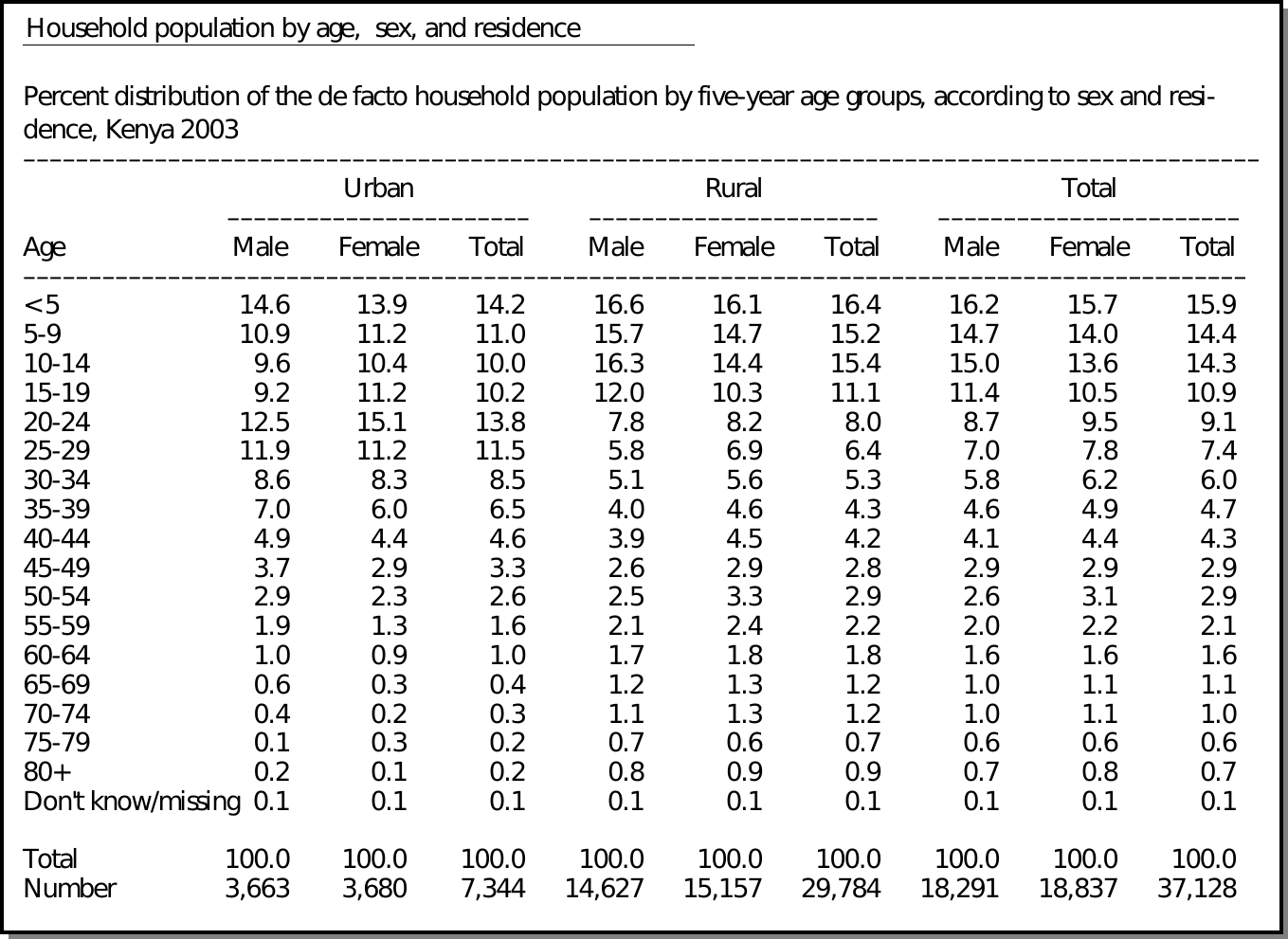}
\caption{Example of available statistics on individuals' characteristics: the Kenyan population by age, gender and residence. Retrieved from \cite[p.~14]{samuel_thiriot:bib_customer_value:kdhs_2003_3}.}
\label{fig:stats_attributes}
\end{figure} 

\begin{figure}[tph]\sidecaption
\includegraphics[width=0.8\textwidth]{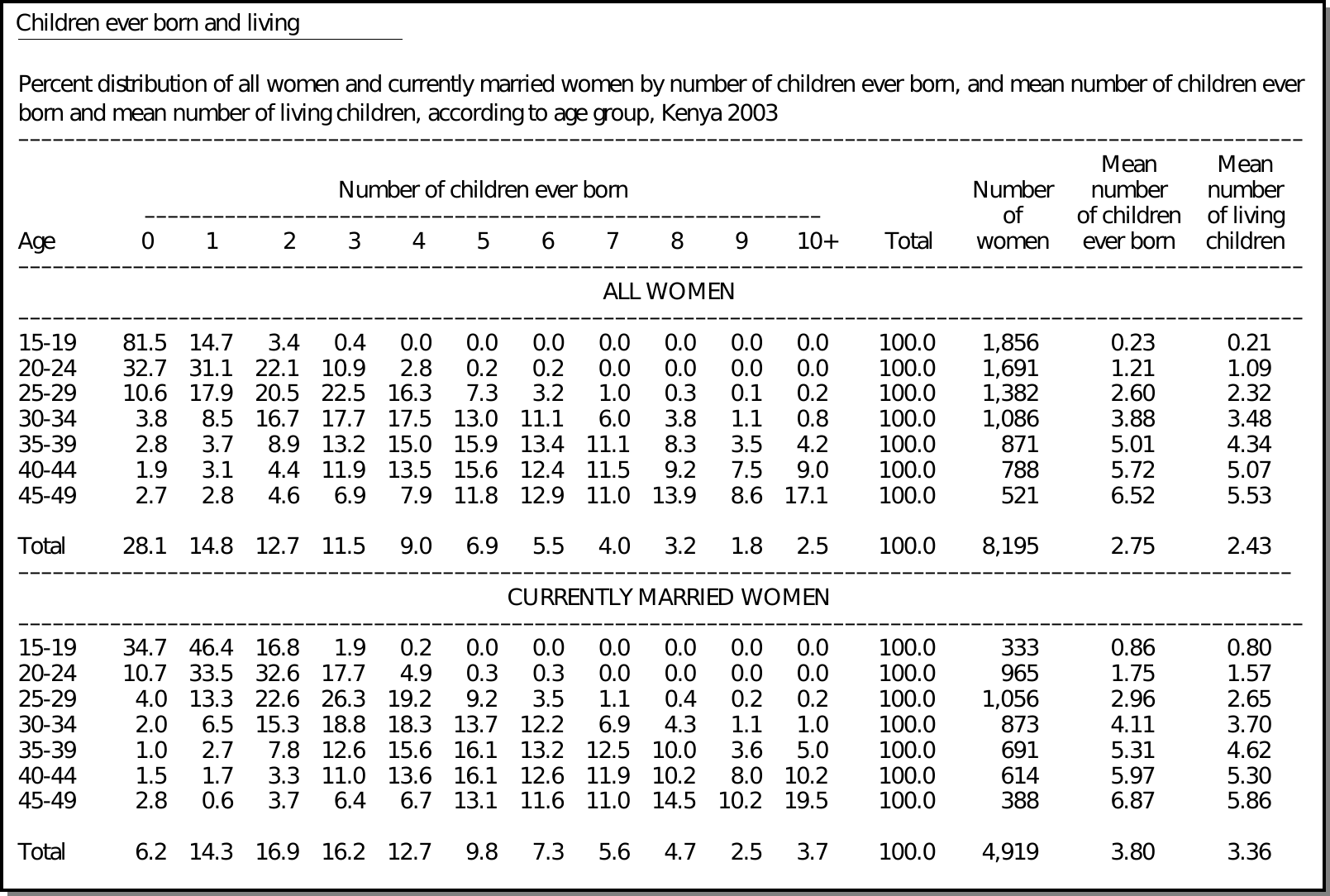}
\caption{Example of available statistics on affiliations given individuals' attributes: school attendance given gender, age slices and residence. Retrieved from \cite[p.~14]{samuel_thiriot:bib_customer_value:kdhs_2003_3}.}
\label{fig:statsPopu}
\end{figure}

% As said previously, the social structure is assumed to be not collectable. The lack of observations on real large networks is precisely considered to be the main obstacle to network reconstruction. However, many observations exist for any population, at a country or regional scale.As said previously, the 

As said previously, many observations on the structure of interaction exist for any population. These indirect observations \citep{samuel_thiriot:bib_psycho:yin_2003_1} come from many sources: National Census, published sociological studies on the same population, public statistics from national institutes of statistics and qualitative feedback from field experts. These observations cover several aspects that match the social processes cited before.

% statistics // attributes
National census propose up-to-date and statistically representative data on the \textit{characteristics of individuals}. These statistics include information on age, socioeconomic classes, gender, religion, type of habitat (rural or not) and other characteristics. National census are available for most countries, and are published every 3 to 5 years. Other national or regional statistics are often also available. Socioeconomic characteristics are often interdependent. They are often published as statistical correlations depending to other characteristics. For instance, educational level depends on age and gender. An example of statistical tables that are published is provided in figure~\ref{fig:stats_attributes}. These statistics may be used to reproduce a population of agents in which characteristics will reflects those of the real population. \textit{These attributes are related to homophily, and will be used as a basis to create links given individuals' characteristics. }

% statistics // number of links
Evidence is also available, with more or less precision, for the \textit{number of relationships of various kinds} that people possess. Family links are the more documented, because demographics is part of national census. The number of children per man or woman (given age, gender and educational status) is often measured and published (see Fig.~\ref{fig:example_data_kenya_maritalstatus} p~\pageref{fig:example_data_kenya_maritalstatus}), as well as the marital status. These statistics National Census and sociological studies inform us on the \textit{opportunities people have to meet}. Affiliations of people given their attributes are often quantified, like the number of children attending schools, being involved in a sport club, going church or being employed (e.g. Fig.~\ref{fig:statsPopu}). This data may be used to parameter baseline homophily.

% statistics // frequency or nature of interactions

% In fact, no generic study can be used on interactions. The adequate information have to be selected given the social phenomenon modelled. If the modeller is interested in word-of-mouth, s/he will focus its search on the frequency of interaction in the workplace, in family and among friends (e.g. \citep{samuel_thiriot:bib_customer_value:carl_2006_1}). If s/he studies TODO...

% The number of links leading to interactions of different natures is either studied at country or local scale. 
% For instance, the French National Institute of Statistics (INSEE) publishes ponctual studies on the frequency of interactions with colleagues, neighboors and son on. 
% conclusion

\textit{These statistics are scattered}, in the sense that they come from various sources and where collected at different scales. They have to be completed by qualitative observations from field experts. However, \textit{they constitute the only available data on the characteristics of the population and on the structure of relationships}. Rather than relying on statistics from other networks to generate the network, \textit{we propose to rely on that field data to generate a network adapted to both the local culture and the social phenomenon studied by the user}. Moreover, these observations are strongly related to the two main generative rules introduced before, and could be used to parameter such types of rules.

\begin{framed}
In the case of Kenya, a huge lot of statistics are provided by the Kenya Demographic and Health Survey (KDHS) like age, gender, religion, employment, attitude against contraceptive use or number of children per women in age to procreate \citep{samuel_thiriot:bib_customer_value:kdhs_2003_3}. Sociological studies detail the structure of families in Kenya \citep{samuel_thiriot:bib_customer_value:mburugu_2004_1}. 
Field studies on diffusion of contraceptive use constitute a valuable source of qualitative observation \citep{samuel_thiriot:bib_customer_value:watkins_1995_2,samuel_thiriot:bib_customer_value:rutenberg_1997_1}.
\end{framed}

\section{User methodology\label{indoc:user_methodo}}

\subsection*{Outline}

As we build a tunable generator that may generate various types of networks (relationships, family structure, etc.) and that accepts observations from the field as parameters, we also propose a methodology that guides users in the formalizing part of the process. This methodology includes three main steps:
\begin{itemize}
\item \textbf{Step A}: the modeler first chooses the content of the network, that is to define the \textit{attributes} that will be took into consideration for agents and the \textit{types of social links} that will exist in the multiplex network.
\item \textbf{Step B}: the user then describes the \textit{interdependencies between attributes} from scattered statistics and qualitative observations.
\item \textbf{Step C}: the modeler then defines \textit{the generation rules} that will actually be applied by the generator. S/he provides \textit{the probabilities to link agents given their attributes}, \textit{transitivity rules} and the order in which these rules will be applied.
\end{itemize}

\begin{table}[thp]\sidecaption
\includegraphics[width=0.7\textwidth]{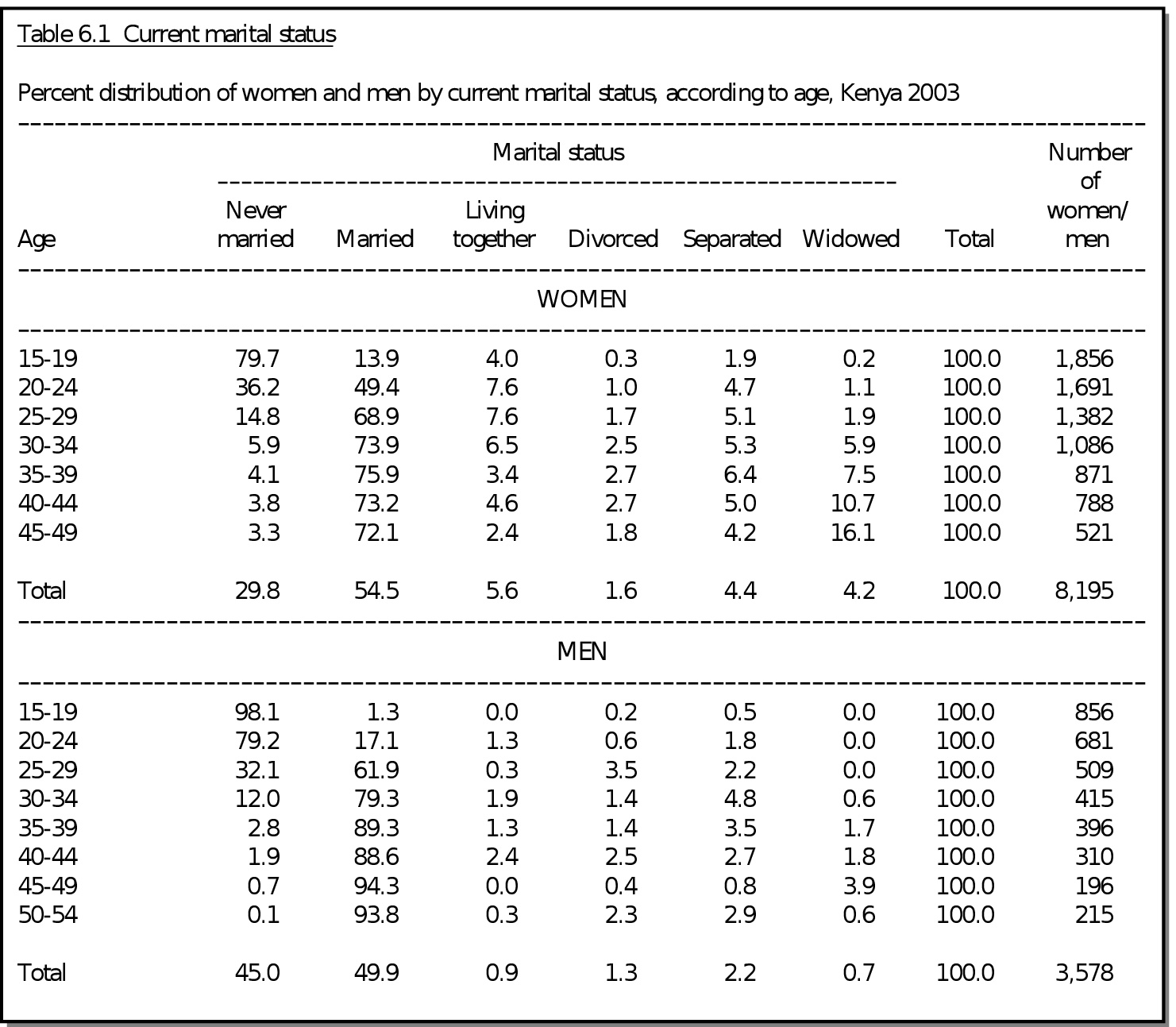}
\caption{Example of statistical data available for individuals' attributes: number of wives per men given age slices (data extracted from \citep[p.~91]{samuel_thiriot:bib_customer_value:kdhs_2003_3})}
\label{fig:example_data_kenya_maritalstatus}
\end{table}

\subsection{Step A: determine the content of the network}

\subsubsection{Choose the kinds of social links $\mathcal{T}$}

As said previously, we will generate multiplex networks that will include various types of links. Before anything else, the modeler has to \textit{define the possible types of links}, by defining the set $\mathcal{T}$. For instance, a modeler may choose $\mathcal{T} = \{t^{married},t^{children},t^{neighboors}\}$.

The reasons for adding a kind of social link are:
\begin{itemize}
\item \textit{The nature of this social link leads to different interactions}, either in a quantitative (frequency of the interaction) or qualitative (social influence) way. For instance, when modelling word of mouth, strong and weak ties lead to different persuasion powers and different frequency of information \citep{samuel_thiriot:bib_customer_value:carl_2006_1}
\item \textit{The nature of social link has to be differentiated from others during generation}. As example, if we aim to describe the fact that a student has more probabilities to bond with friends of the same college than with friends of its original birthplace, we have to differentiate friendship links in college and friendship links from childhood. 
\end{itemize}

The modeler also has to precise, for each kind of link, if these links are directed or undirected. The principle is to use undirected links by default, and directed links only when it is mandatory in the simulation or for the generation process. 

\begin{definition}
The set of possible network types $\mathcal{T}$ defines the various kinds of social links that can be described into the generated network. $\mathcal{T}^{dir}$ and $\mathcal{T}^{undir}$ (with $\mathcal{T} = \mathcal{T}^{dir} \cup \mathcal{T}^{undir}$) define respectively the sets of directed and undirected types of social links.
\end{definition}

% illustration Kenya
\begin{framed}
In our illustration in Kenya, observations from the field indicate the kind of relationships that lead to different interactions on contraceptive solutions. 

Gossip on the possible risks on contraception mainly take place in the workplace, when women work in fields or men at the market \citep{samuel_thiriot:bib_customer_value:rutenberg_1997_1}. Contraceptive solutions are considered to be a controversial topic, so women prefer to discuss it with their friends of the same age, rather than with family-in-law. Family links lead to normative pressure against adoption: mothers don't like to see their daughters adopt a contraceptive, and men are considered by their father to be too soft if they allow their wife to adopt. However, normative pressure may also be positive, as between siblings, with men or women adopting more easily if one sibling has adopted. Final adoption of contraceptive solution has to be negotiated between the wife and her husband. 

Given these observations, we have to differentiate links between wives and their husbands (named ``spouses''), friendship relationships (``friendship''), siblings (``siblings''), colleagues and parenthood. As will be shown later, we have to split parenthood links for making the generation possible. We then add motherhood links (``motherOf'') and fatherhood (``fatherOf'') links. Here $\mathcal{T}=\{\mathit{spouses, motherOf, friendship, siblings, fatherOf}\}$.

While friendship or neighborhood don't require directivity, motherhood links (``mother of'') have to be directed both for simulation and construction of the network. The social influence process against adoption is directed from mothers to their daughters and not the opposite way. Also, to create siblings, we need to identify the children of the same mother, which is impossible if the link is not directed. Thus $\mathcal{T}^{dir}=\{\mathit{motherOf, fatherOf}\}$ and $\mathcal{T}^{undir}=\{\mathit{spouses, friendship, siblings, colleagues}\}$.
\end{framed}

% It appears that men or women discuss mainly with people of the same gender. They also often discuss when working with their colleagues. Works are clearly different given gender: women retrieve water from the river and work in the fields, while men meet at the market. 

\subsubsection{Choose attributes to describe $Att$}

The modeler then chooses the characteristics that will defined for agents in the heterogeneous population $\mathcal{A}$. These characteristics are named ``attributes'', and are noted $Att$. Each attribute of an agent may take a value in the discrete set of possible values (named the \textit{domain}) for each attribute. At this step, \textit{the modeler should define both the attributes and the domain of these attributes} (the domain is the set of possible values for each attribute).

There are three main reasons that justify the add of an attribute into the network:
\begin{itemize}
\item The attribute is assumed to influence the existence of social links.
\item The attribute is required to compute other attributes. 
\item The attribute will be used during the simulation to change the individual behavior.
\end{itemize}

According to these reasons, \textit{attributes include both affiliations and degree of connectivity}. Affiliations influence the creation of social links; in fact, being linked to the same workplace may be sawn as a special case of homophily. Degree of connectivity is rarely required in models, being rather asked to the modeler as a density parameter \citep{samuel_thiriot:bib_psycho:robins_2001_1}. However, as discussed previously, degree for each kind of link is often available from field observations. When it is not, it is easier to collect information, or define hypothesis, on the number of links of each kind for an individual than estimating the  density of an unknown network. 

\label{indoc:dicrete_domains} Note that we only use discrete domains for variables into Bayesian Networks. In fact, observations are always published in a discrete way.

% illustration Kenya
\begin{framed}
In the case of Kenya, available observations indicate several attributes that determine the existence of links:
\begin{itemize}
\item spouses live in the same \textit{spatial location} (village or town). They have similar \textit{age}, but wives are always several years younger than their husband. We will thus define an attribute ``age'' with discrete values (0..99) and a ``gender'' attribute taking values $\{male,female\}$.  For this illustration, we define abstract ``spatialLocation'': $\{village1, village2, R1, R2... R12\}$.
\item colleagues meet and bond given their main \textit{occupation}. We define two attributes ``workWater'' and ``worksMarket'' with domains $\{yes,no\}$, that represent the fact that an individual may work both in water and market. 
\item friendships often occurs between people of the same age, gender and spatial location. All of these attributes where previously defined.
\item the number of children per mother is available, given her \textit{marital status}  ($\{ married, single \}$) and her age. 
\end{itemize}

The statistics available for Kenya are often presented by age slides (0-14, 15-19 and so on) rather than for each value of age. We then define an attribute ``ageSlices'' with domain $\{ 0-14, 15-19, 20-24 ... 50-54 \}$ that will enable the computation of other attributes from available data. 

During this step, we have defined $Att$ as $\{\mathit{age, gender, spatialLocation, workWater, workMarket, maritalStatus}\}$.
\end{framed}

\subsection{Step B: Describe attributes interdependencies}

Attributes of individuals in a real population are strongly interdependent: marital status depends on age and gender, socioeconomic class is highly correlated with location and education level, etc. Generating a population of agents in which attributes of agents comply with statistical interdependencies of individual characteristics requires a relevant modeling of these dependencies, generic enough to be used with any kind of data. It appears that Data available for a population is often presented as statistics linking one attribute with another. For instance, marital status is provided given age and gender (Fig.~\ref{fig:example_data_kenya_maritalstatus} p.\pageref{fig:example_data_kenya_maritalstatus}). That kind of statistic can be translated, without loss of generality, to conditional probabilities, like $p\left(RC^{a}_{motherOf}=\{0...10\} | age(a), maritalStatus(a)\right)$. In that viewpoint, \textit{attributes of agents can be considered to be random variables}. We propose to \textit{use Bayesian networks to formalize these complex interdependancies between individuals' characteristics}.

\begin{figure}[ht]
\centering
\includegraphics{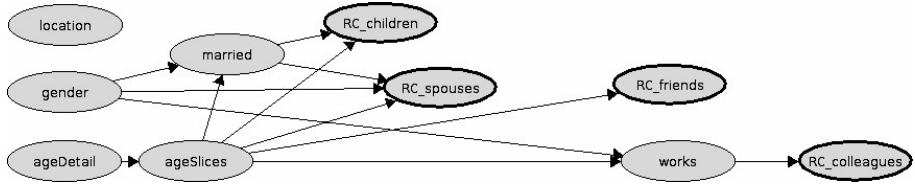}
\caption{Attributes Bayesian network used to describe interdependencies between Kenyan socio-demographic attributes. Nodes in bold are the number of links to create for each link type.}
\label{fig:attributesBN}
\end{figure} 

A Bayesian Network (or directed acyclic graphical model) \citep{samuel_thiriot:bib_model_decision:jensen_1996_1}, later named BN, includes random variables $V=\{V_1...V_i\}$ and their conditional dependencies using a directed acyclic graph. Nodes $\{V_1...V_i\}$ in the network correspond to random variables. Each variable may take various values, named its domain $D^i$. As said before, we will only use discrete domains. Links traduce statistical dependencies between the random variables. The absence of link between two variables means these variables are assumed independent. Bayesian Network provide a compact representation of a probability density in a possibly large space of variables. Note that creating a Bayesian Network is far from abstract for users, as they can use free software that enables to edit them with a graphical user interface (for instance SamIam).
% TODO capture écran ?

Each agent attribute in $Att$ is represented by a variable in the BN. The domain of each variable defines the values the attribute can take. For instance in graph~\ref{fig:attributesBN}, variable \textit{gender} has domain $D^{gender}=\{ male, female \}$, $D^{married}=\{yes,no\}$, and $D^{RC\_motherOf}=\{0...10\}$. Root variables define initial probabilities. In Fig~\ref{fig:attributesBN}, initial probabilities for variable \textit{ageDetail} define the probability for an individual picked up randomly in the population to have a given age; that probability is available from the age pyramid of the target population. A directed link between two variables $V_1 \rightarrow V_2$ means that $V_2$ probabilities can be computed using its parents, and only its parents. $V_2$ embodies a conditional probability table representing the probability to take each value $D^{V_2}$ given all the possible values in the domains of its parents (here, $V_1$). No link means that variables are assumed independent. The absence of a link don't means that variables are independent in reality, but rather reflect our lack of knowledge (or our willingness to simplify that knowledge) of that dependence.

\begin{definition}
The \textit{Attribute Bayesian Network} (Attribute BN) is a Bayesian Network passed to the generator as a parameter. It describes the attributes of agents corresponding to individual's characteristics, their possible discrete values for each attribute and the statistical interdependancies between attributes. It is used by the generator for creating a population of heterogeneous agents whom attributes are statistically plausible.
\end{definition}

\begin{framed}
In our application to Kenya, probabilities in the agent BN depicted in Fig.~\ref{fig:attributesBN} come from the US Census Bureau, from the Kenya demographic and health survey \citep{samuel_thiriot:bib_customer_value:kdhs_2003_3}, and from field studies (e.g. \citep{samuel_thiriot:bib_customer_value:watkins_1995_2,samuel_thiriot:bib_customer_value:rutenberg_1997_1}).

Initial probabilities for attribute \textit{ageDetail} directly come from the age pyramid of Kenya (see~\ref{fig:stats_attributes} p.~\pageref{fig:stats_attributes}). The age pyramid of Kenya is symmetrical, so the initial probabilities of variable \textit{gender} are identical ($p(gender=male)=p(gender=female)=0.5)$. As discussed when defining the variables' domains, age is often published as slices. We then define the variable \textit{ageSlices} as making an approximate mapping from the discrete age \textit{ageDetail}. Its conditional probability table is defined such as $\forall D^{ageDetail} \in [0:14]$, $p(ageSlice='1-14')=1$, $\forall D^{ageDetail} \in [15:19]$, $p(ageSlice='15-19')=1$, and so on. 

% TODO the figure don't includes waterEau 

The conditional probability table of variable \textit{maritalStatus}, shown in Tab.~\ref{tab:example_cpt_maritalstatus}, is translated from the statistics presented in table~\ref{fig:example_data_kenya_maritalstatus}. Probabilities for variables \textit{works}, or the number of links children and spouses are build in the same way. No statistics are available for the number of friends nor the number of colleagues; we chosen them according to qualitative observations in the field studies.
\end{framed}

\begin{figure}[t]\sidecaption
\begin{tabular}{|c|r|r|l|l|l|l|}
\hline
\multicolumn{2}{|c|}{\textbf{gender}}		& \multicolumn{2}{|c|}{male} & \multicolumn{2}{|c|}{female}\\
\hline
\multicolumn{2}{|c|}{\textbf{married}}	& no	& yes 	& no	& yes	\\
\hline
	& 0-14		& 1.0	& 0.0	& 1.0	& 0.0 	\\
	& 15-19		& 0.981	& 0.019 & 0.506	& 0.494 \\
	& 20-24		& 0.816	& 0.184 & 0.311 & 0.689 \\
\textbf{age}	& 25-29		& 0.381	& 0.619 & 0.261	& 0.739 \\
\textbf{slices}	& 30-34		& 0.207	& 0.793 & 0.241	& 0.759 \\
	& 35-39		& 0.107	& 0.893 & 0.268 & 0.732 \\
	& 40-44		& 0.114	& 0.886 & 0.279 & 0.721 \\
	& 45-49		& 0.057	& 0.943 & 0.279 & 0.721 \\
	& 50-54		& 0.062	& 0.938 & 0.279 & 0.721 \\
\hline
\end{tabular} 
\caption{Example of conditional probability table for variable maritalStatus ($p(maritalStatus | gender, age)$) build from the statistics shown in Fig.~\ref{fig:example_data_kenya_maritalstatus} p~\pageref{fig:example_data_kenya_maritalstatus}}
\label{tab:example_cpt_maritalstatus}
\end{figure}

% 
% Another benefit of BN is to \textit{highlight evident discrepancies in data}. For instance, a social scientist will immediately note in (Fig.~\ref{fig:bayesian_network_1}) the absence of link between gender and age, while the age pyramid in most of countries shows significant differences between genders (indeed, Kenya is a particular case of symmetrical age pyramid). 

\subsection{Step C: Parameter the generation rules}

\subsubsection{Choose the generation process}

The modeler can now \textit{describe which rules will be applied for generating the network}. Given his/er observations from the field and/or his/er hypothesis on the reasons for a link to appear in the network, the modeler will describe the successive steps that will be used by the generator to create the links. These steps constitute the generation process, based on a set of generative rules $\mathcal{R} = \{ R_1...R_i \}$.

As explained before (see~\ref{indoc:principles}), two kinds of rules may be defined: creation of a link by homophily or by transitivity. The intuitive meaning of these rules is:
\begin{itemize}
\item Attributes: ``Link $t$ is created given the agents attributes.'' For instance, the social relationship ``spouses'' is only plausible if we take into account at least age, gender and spatial location of agents. 
\item Transitivity: ``Link $t$ may appear because of the existence of two other links with a third acquaintances''. 
\end{itemize}

Note that a kind of link $t \in \mathcal{T}$ may be created by several generation rules. A modeler can then first define friendship as depending of common characteristics (attributes), but being also created by transitivity accross previous friendship links. It is important to underline that \textit{the order of rule creation will have a considerable impact on the generated network}. This is principally due to the impossibility to create multiple links between individuals. The order of the rules enables the modeler to prioritize the relationships, and to eliminate implausible relationships. 

\begin{definition}
The \textit{ordered} set $\mathcal{R}$ defines the generative rules that will be applied for actually generating the network. Rules $R \in \mathcal{R}$ may be either homophily or transitivity rules. The sets defined such as $\mathcal{R} = \mathcal{R}^{homophily} \cup \mathcal{R}^{transitivity}$ may be used to determine if a given generative rule $R \in \mathcal{R}$ uses homophily ($R \in \mathcal{R}^{homophily}$) or by transitivity.
\end{definition}

% 
% For instance, once a ``spouses'' link created between two agents, it becomes impossible to add a 
% 
% The order of generatives rules in $\mathcal{A}$ thus reflects two thinks: 
% \begin{itemize}
% \item a kind of priorization of the nature of links. In the simulation, two people known to be spouses aren't considered to be colleagues; 
% \item the representation of social rules: a man cannot marry his daughter and siblings neither.
% \end{itemize}

\begin{framed}
In the case of Kenya, we parameter the generator in order to successively create:
\begin{itemize}
\item spouses links from attributes, in order to link men and women of compatible age and spatial location
\item motherOf links from attributes, by linking each mother to the required number of children of compatible age
\item fatherOf links by transitivity, in order to close tryads ``spouse'' and ``motherOf''. This enables us to create links between fathers with the children they had with one of their wives. Once a fatherOf link created, its existence forbids to create another kind of link like colleagues or friendship.
\item siblings by transitivity between children sharing the same mother or the same father. 
\item friendship from attributes, between people having the same age and often the same location
\item colleagues from attributes, between people living in the the same location and working on the same activity
\end{itemize}
\end{framed}

\subsubsection{Represent homophily using matching BNs}

As done in SNA, we consider the existence network links as random variables. The existence of a link of type $t$ between agents $a_1$ and $a_2$ is written $L^t_{a_1,a_2}=1$, $L^t_{a_1,a_2}=0$ meaning that no link exists between the agents. A probability law $p(L^t_{a_1,a_2}=0)$ may be written (see table~\ref{tab:notations} p.~\pageref{tab:notations} for all the notations used within the paper). 

We previously defined generation rules that create links given attributes of individuals. For each of these rules, we would like to describe the probability of link existence given the attributes of agents. Such a rule may be written $p(L^{t_i}_{a_1,a_2}=1 | Att(a_1), Att(a_2))$, which is a conditional probability that depends on the values of agents' attributes. As example, a link representing motherhood would link a woman old enough for procreating with a younger individual, both living in the same spatial location if the child is young enough.  

\begin{figure}[t]
\centering
\includegraphics{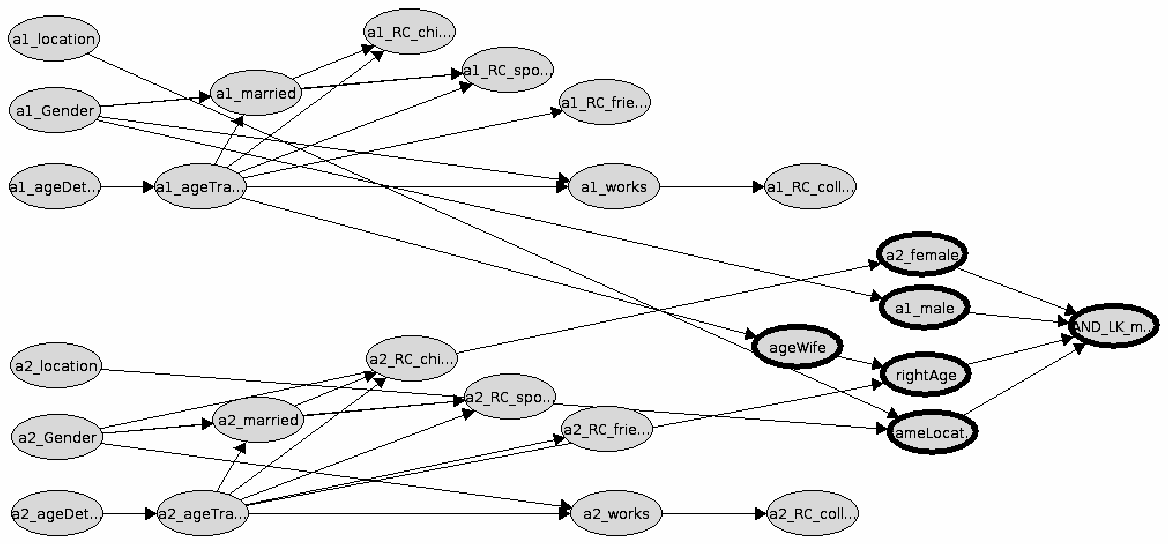}
\caption{Matching Bayesian network for link type \textit{spouses} for the illustration of Kenya. On the left, the agent BN for agents 1 and 2.}
\label{fig:bn_appariement_1}
\end{figure}

We also propose to represent these probabilities with Bayesian Networks. We name ''matching Bayesian Networks`` (or matching BN) a Bayesian Network that encodes this probability law. An example of such a matching BN is provided in figure~\ref{fig:bn_appariement_1}. A matching BN always includes:
\begin{itemize}
\item The random variable $L^{t_i}_{a_1,a_2}$. In the matching BN, this variable is formalized with a Boolean variable (that is, a variable having for domain $\{yes,no\}$). In figure~\ref{fig:bn_appariement_1}, this variable is shown at the right of the graph.
\item Nodes representing the attributes of agents $a_1$ and $a_2$. In figure~\ref{fig:bn_appariement_1}, the nodes on the left prefixed by ``a1'' or ``a2'' correspond respectively to the attributes of two agents $a_1$ and $a_2$ belonging to $\mathcal{A}$. These nodes are noted $\mathcal{V}^{Att1}$ and $\mathcal{V}^{Att2}$. $\mathcal{V}^{Att1} \subset \mathcal{V}^{Att}$: all nodes are not necessarily used for matching, but they all should have been defined in the attributes BN first. 
\item Nodes that compute conditions on matching. These nodes are drawn in bold in this figure. They take as parents the attributes of agents $a_1$ and $a_2$ and have the  random variable $L^{t_i}_{a_1,a_2}$ as a child.
\end{itemize}

Note that matching BN offer a great flexibility to the user. One can define assortativity by defining an higher probability to create link between people having both an high degree, or disassortativity by using the opposite principle. On the contrary, preferential attachment would simply by defined in the attributes of agents, in the node describing the required degree, with an high probability to have few links and a low probability to have an higher degree.

\begin{definition}
A \textit{matching Bayesian Network} (matching BN) is a Bayesian Network passed to the generator as a parameter. It describes the conditions of creation of links of a given type $t \in \mathcal{T}$, depending to the values of the attributes of agents. It may be used to describe baseline and inbreed homophily, assortativity or disassortativity or affiliations. A matching BN always includes a boolean variable that computes the probability to create (or not) a link between two agents given the values of their attributes. This variable may be exprimed as $p(L_{a_1,a_2}^t=1 | Att(a_1), Att(a_2))$. One matching BN per attribute generation rule $R \in \mathcal{R}^{homophily}$ has to be provided to the generator.
\end{definition}

% Note that at this step, the probability of the node TODO reflects the probability for two agents picked up randomnly in the population, and not being TODO, to be linked by a social link of type $t_i$. 

\begin{framed}
The creation of such a network is quite intuitive. Let us take the example of spouses in Kenya. We start from a skeleton matching BN (that is, in fact, generated by our software). This skeleton includes the mandatory variable $L^{t_i}_{a_1,a_2}$ (on the right), and the attributes of agents on the left (which are just copies of nodes from the Attribute BN prefixed by either ``a\_1'' or ``a\_2'').

The matching BN in Fig.~\ref{fig:bn_appariement_1} represents the conditions used for the creation of link ``spouses'' in Kenya. We define arbitrarily that agent 1 is male and agent 2 female (wedding in Kenya is always heterosexual). Node \textit{ageWife} projects the probable age of the first wife of man described on top (on average 10 years younger), and variable \textit{rightAge} ensures by an identity probability table that agent 2 complies with that age. The node \textit{sameLocation} takes value \textit{yes} only if both agents live in the same location. The final variable ``linkSpouses'', which determines if two agents can be linked together, takes values ``yes'' only if all of its parents are themselves to ``yes''. Note the nodes \textit{a1\_created\_spouses} and \textit{a1\_remaining\_spouses}, which ensure that we will only create as many links of type $t$ as required by $RC_t^{a_1}$ and $RC_t^{a_2}$, but no more, so a wife will exactly be tied with one husband (these nodes and the relevant conditional probability tables are created and managed by the generator). 

Other links in $\mathcal{T}^{Att}$ are defined in the same way: friends have probably the same age and live probably in the same town, mothers are linked to children whom age is compliant with their age (and live always in the same location if children are young), and colleagues are defined as agent sharing the same activity in the same location (see table~\ref{tab:rulesKenya} p~\ref{tab:rulesKenya}).
\end{framed}

\subsubsection{Describe transitivity rules}

\begin{definition}
A transitive rule has to be detailed for each $R \in \mathcal{R}^{transitivity}$. A \textit{transitivity rule} describes the probability $p(L_{a_1,a_3}^{t_3}=1)$ to create a link of a given kind $t_3 \in \mathcal{T}$ between two agents $a_1,a_2 \in \mathcal{A}$, given the fact that $a_1$ and $a_3$ share a common acquaintance $a_2 \in \mathcal{A}$ that is linked with $a_1$ and $a_3$ with links of a given kind $t_1,t_2 \in \mathcal{T}$. It may be exprimed as $p(L_{a_1,a_3}^{t_3}=1 | L_{a_1,a_2}^{t_1}=1, L_{a_2,a_3}^{t_2}=1)$ with $a_1,a_2,a_3 \in \mathcal{A}$, $a_1 \neq a_2 \neq a_3$ and $t_1, t_2, t_3 \in \mathcal{T}$. 
\end{definition}

\begin{framed}
For our application in Kenya, we defined $\mathcal{R}^{transitivity} = \{ \mathit{fatherOf}\}$. Transitivity enables us to ensure that an agent will be said ``father of'' the same children than the children of his wives. We want to create a link between each dyad of agents $A1$ and $A3$ for agents $A1$ married with $A2$, with $A2$ mother of $A3$. Such a rule is written $p(L^{fatherOf}_{a_1,a_3}=1 | L^{spouse}_{a_1,a_2}=1, L^{motherOf}_{a_2,a_3}=1)=1$. We set this probability to 1, because culture in this area of Kenya don't allows women to broke weddings. In a more European culture, we would have set a lower probability, and would have added other rules to create links with the ex-wife of divorced men that would have been use to create paternity relationships. The generator will process the possible combinations of $a_1$, $a_2$ and $a_3$ in such a rule given the directivity of links defined in $\mathcal{T}^{dir}$ and $\mathcal{T}^{undir}$.
\end{framed}

\subsection{Parameter the generator}

\begin{table}[t]
\begin{tabular*}{\textwidth}{|p{0.25\textwidth}|p{0.68\textwidth}|}
\hline
$N$				& size of the population to generate \\
$\mathcal{T} = \{ t_1...t_M \}$ & the kinds of social links that may exist in the multiplex network \\
$\mathcal{T} = \mathcal{T}^{dir} \cup \mathcal{T}^{undir}$ & defines which links are directed and undirected in $\mathcal{T}$ \\
\hline
Attribute BN			& the attribute Bayesian Network that describes the attributes of agents, their domains and their interdependancies \\
\hline
$\mathcal{R}$			& the generations rules (ordered) that will be applied to generate the population \\
$\mathcal{R} = \mathcal{R}^{homophily} \cup \mathcal{R}^{transitivity}$	& defines which rules are created by transitivity or homophily\\
Matching BN			& $\forall R \in \mathcal{R}^{homophily}$, the Bayesian Network that describes conditions on linking \\
Transitive Rules		& $\forall R \in \mathcal{R}^{transitivity}$, the rule of generation of transitive links \\
\hline
\end{tabular*} 
\caption{Parameters of the generator}
\label{tab:parameters}
\end{table}

\begin{table}
\begin{tabular*}{\textwidth}{|l|p{0.8\textwidth}|}
\hline
method & conditions \\
\hline
homophily  & ``motherOf'' links are created between A1 and A2, with A1 women and A2 being younger than A1. When A1 is younger than 15, s/he has to live in the same spatial area than A1.\\
homophily & ``Spouses'' links are created between A1 and A2, with A1 a man and A2 a women living in the same place. \\
transitivity & ``FatherOf'' links are created by transitivity between A1, A2 and A3 with probability 1, when A1 is the husband of A2 and A2 motherOf A3. \\
homophily   & ``friendship'' links are created mainly between agents having the same age (most of the time) and living \textit{often} in the same location \\
homophily  & ``colleagues'' links are created between people working at the same activity and living in the same place \\
\hline
\end{tabular*}
\caption{Rules of link creation for the example of Kenya, as could be summarized in a publication of a model of innovation diffusion.}
\label{tab:rulesKenya}
\end{table} 

This methodology led to the formalization of all the required parameters required by the generator (listed in table~\ref{tab:parameters}). For the example of Kenya, generative rules are defined as summarized in \ref{tab:rulesKenya}. 

% The current approach is to generate a new network at the beginning of each simulation. This online use of generators is possible for cost-less networks generator. However, generating more descriptive networks becomes more costly, and would involve some TODO. We propose to generate a large number of networks before the actual simulation, and to store them in a networks library. When enough networks are stored, they already provide a good exploration of the possible networks. 

\section{Generator: algorithms and implementation\label{indoc:generation}}

\subsection{Outline of the algorithm}

The generation of the population is done according to the following algorithm. The algorithms for matching agents given their attributes and creating links by transitivity are detailed below.

\begin{algorithm}{H}
\begin{algorithmic}[1]
\small
\REQUIRE $N$ the population size
\medskip
\STATE load parameters $\mathcal{R}$, $\mathcal{T}$, $Att(a)$
\STATE create database for agents and network storage given $Att(a)$ 
\STATE generate the population of $N$ agents
\FOR{each $R \in \mathcal{R}$ }
	\IF{$R \in \mathcal{R}^{homophily}$}
	\STATE use the attributes matching algorithm provided along $R$ to create the links
	\ELSE 
	\STATE // $R \in \mathcal{R}^{transitivity}$
	\STATE use the algorithm for creating transitivity links given the $R$ rule
	\ENDIF
\ENDFOR
\STATE export the population $\mathcal{A}$ with $\forall a \in \mathcal{A}, Att(a)$ into a file
\STATE export the network  $X(A,L)$ 
\STATE compute statistics on the network 
\end{algorithmic}
\caption{Main algorithm of the generator}
\label{alg:generatorMain}
\end{algorithm}

\subsection{Algorithm for the generation of agents\label{indoc:generation_proto}}

Our purpose is to generate a population of agents, in which the distribution of attributes' values will comply with the one of the Attributes BN provided as parameter. 

All the variables $Att$ in the agent BN will become agents' attributes with the same domain. For each agent $a \in \mathcal{A}$ to create, we generate a \textit{prototype agent}. The process to generate a prototype simply consists in using the agent BN in a generative way: for each variable $V \in Att$ of the agent BN (in the ordinal order, so root variables are processed first), a value $V=v$ is selected randomly in the domain of $V$, given probabilities $p(V=v | parents(V))$ defined in the BN. When value $V=v$ has been chosen, a corresponding piece of evidence $p(V=v)=1$ is put in the BN. Evidence, in the theory of BN, represents a known information \citep{samuel_thiriot:bib_model_decision:jensen_1996_1}. Putting evidence in the BN permits to compute probabilities of child variables given the values of already selected attributes, so the integrity of agent attributes is ensured. An inference engine is in charge of computing probabilities given evidence.

\begin{figure}[t]
\centering
\begin{tabular*}{12cm}{c}
\vspace{0.5em}\includegraphics{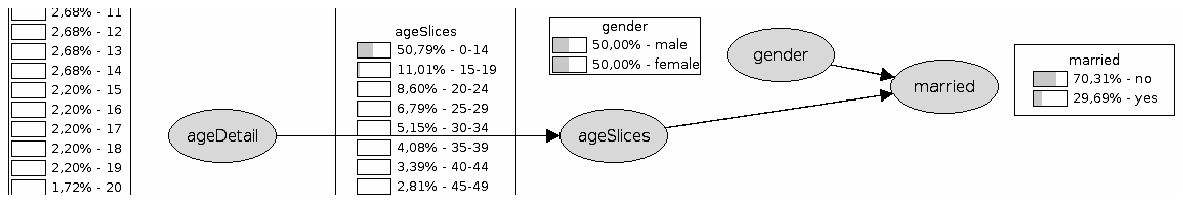} \\ 
\includegraphics{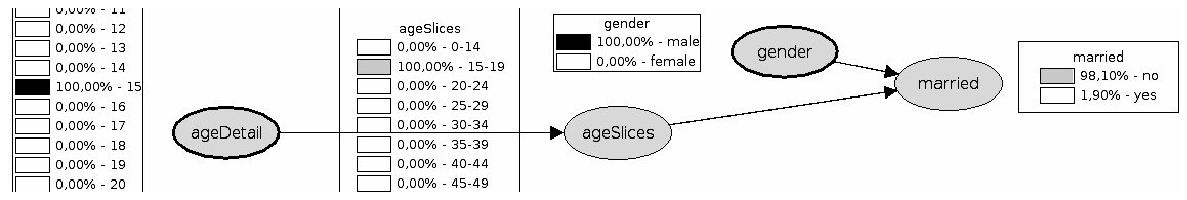}
\end{tabular*}
\caption{\footnotesize Example of evidence propagation when the bayesian network is used to generate agents' attributes. Here monitors (boxes in the figure) display the probabilities for each variable to take every value (note that some of these monitors are truncated). \textit{(top)} probabilities with no evidence \textit{(bottom)} probabilities when evidence is set.}
\label{fig:evidence_propagation}
\end{figure}

For instance in Fig.~\ref{fig:evidence_propagation}, before any piece of evidence \textit{(top)}, the probability for someone randomly picked up in the population to be married is 29.69\%. When attributes $ageDetail$  and $gender$ have been randomly put to $15$ and $male$, and used as evidence, posterior probability for the current agent to be married falls to 1.90\%. When all the agents are generated that way, the statistical distribution of their attributes complies with the distribution described by the BN.

Once the agent $a_1$ generated, that is that $Att(a_i)$ is defined, the agent is added to the population $\mathcal{A}$. The population is stored in a database during the generation, in order to facilitate further set operations on the population.

\begin{table}[t]
\begin{tabular*}{\textwidth}{|p{0.25\textwidth}|p{0.68\textwidth}|}
\hline
\textbf{notation}		& \textbf{description} \\
\hline
$\mathcal{A} = \{ a_1...a_N \}$	& the population of agents \\
$Att(a)$ with $a \in \mathcal{A}$ & the attributes of an agent and their values \\
$a_1 \neq a_2$			& $a_1$ and $a_2$ are no the same agent. \\
\hline
$\mathcal{X}$			& the space of all the possible networks \\
\hline
$\mathcal{T} = \{ t_1...t_M \}$ & the kinds of social links that may exist in the multiplex network \\
\hline
$L_{a_1,a_2}^t=1$		& agents $a_1$ and $a_2$ are linked together with a link of kind $t \in \mathcal{T}$ \\
$p(L_{a_1,a_2}^t=1)$		& probability for two agents to be linked together \\
\hline

$\mathcal{V}=\{V^1...V^n\}$ 	& the random variables of a Bayesian Network \\
$parents(V^i)$			& parents of a variables into a Bayesian Network \\
$V^i=\{v_1...v_p\}$, with $V^i \in \mathcal{V}$	& domain of a variable in a BN, that describe the possible values of this variable \\
$\forall v \in V^i, p(V^i=v) \in [0:1]$ & probabilities described by the BN for each variable, that may depend on other variables ($p(V^i=v | parents(V^i)$ ) \\
$Ev$				& the set of evidence added to the BN for describing peaces of information, that includes sure probabilities for values of various variables (for instance: $Ev= \{ p(V^2=v_3)=1, p(V^4=v_1)=1, p(V^x=v_y)=1 \}$ \\
$\forall v \in V^i, p(V^i=v | Ev) \in [0:1]$ & probabilities described by the BN for each variable, as computed by an inference engine that takes into account evidence $Ev$ on some variables \\
\hline
\end{tabular*} 
\caption{Table of notations}
\label{tab:notations}
\end{table} 

\subsection{Algorithm for matching agents}

Now that all agents were created in the population $\mathcal{A}$, each agent having its attributes $Att(a)$ defined, we have to link them using the matching BN. Building such a matching algorithm faces two challenges. The first is to \textit{create links that comply with the matching BNs} provided as parameters. The second challenge is to \textit{limit the complexity of the algorithm}. The worst solution is to iterate accross all the possible dyads. This solution would be linear with the number of possible links in the network, which is itself combinatorial on the number of agents in the network (the number of possible links in an undirected network of size $n$ is $n(n-1)/2$). 
% We restrict this complexity by several solutions than will be detailed in the algorithm, by using Bayesian Networks in a unusual way that restricts 

% TODO figure with sets

\subsubsection{Sets of candidates agents}

\begin{figure}[ht]\sidecaption
\includegraphics[width=0.6\textwidth]{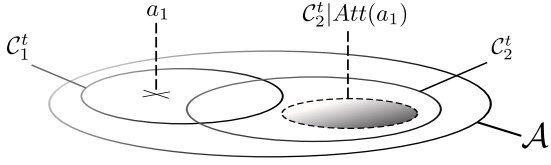}
\caption{The different subsets of the population of agents $\mathcal{A}$ studied during a matching process for a type of link $t \in \mathcal{T}$. $\mathcal{C}_1^t$ and $\mathcal{C}_2^t$ are the sets of agents candidate for linking with link~$t$. The set in gray is $\mathcal{C}_2^t | Att(a_1)$, that is the set of agents $a_2$ candidates to be linked with a given agent $a_1$.}
\label{fig:matching_sets}
\end{figure}

For each kind of relationship $t \in \mathcal{T}^{Att}$, we constraint the matching BN for $t$ by providing \textit{evidence for link creation}: as our aim is to link together agents with link $t$, we set evidence on variable $p(L^{t}_{a_1,a_2}=1)=1$. Given that evidence, probabilities for attributes of $a_1$ and $a_2$ are updated by the inference engine, and some probabilities in attributes' domains fall to zero. For instance, in the case of ``spouses'' link, probability of agents 1 and 2 to be younger than 15 falls to zero; they also cannot have ``maritalStatus=single''. In other words, probabilities in the matching BN given evidence of link creation designate two sets of \textit{candidates for linking} $\mathcal{C}^t_1$ and $\mathcal{C}^t_2$, subsets of the population $\mathcal{A}$ (see Fig.~\ref{fig:matching_sets}). In our application to Kenya, for link type $spouses$, $\mathcal{C}^{spouses}_1$ is the set of husbands and $\mathcal{C}_2^{spouses}$ the set of wives. The matching process will remain limited to these sets, thus limiting the number of links that will be studied.

Then, we iterate across candidates $\mathcal{C}^t_1$.
% and select randomly as many acquaintances among $\mathcal{C}^t_2$ as required by $RC^{a_1}_t$. 
For each agent $a_1 \in \mathcal{C}^t_1$, we load its attributes and use them as pieces of evidence in the BN. After a run of the inference engine, the probabilities for agent 2 define a restricted set $\mathcal{C}^t_2|Att(a_1) \subset \mathcal{C}^t_2$ of candidates for linking \textit{given agent 1 attributes}. In the example of the spouses link in Kenya, once a given $a_1$ chosen, the set $\mathcal{C}_2 | Att(a_1)$ limits $\mathcal{C}_2$ to wives who live in the same location than $a_1$.

\subsubsection{Finding statistically plausible peers}

It is important to underline that $\mathcal{C}_2 | Att(a_1)$ is the set of agents that \textit{may} be linked with $a_1$, but that \textit{all these agents don't have the same probability to be linked} with $a_1$. For instance, given $a_1$ a 44 years old man that live in location $R_3$, $\mathcal{C}_2 | Att(a_1)$ contains all the women in age to be married in the same location, aged from 15 to 45 years. However, it is very unlikely for $a_1$ to be married with a 15 years old girl. He his more probably married with a thirty years old wife. The set $\mathcal{C}_2 | Att(a_1)$ contains all the possible candidates, even if they have a very small probability to be selected. 
% The set $\mathcal{C}_2 | Att(a_1)$ may also (rarely) contain agents that cannot be plausibly linked with $a_1$, because of complex conditions on matching that somewere encode a ``or'' operation. As example, 
We now have to use the probabilities encoded in the matching BN for the actual matching.

Iterating accross the whole set $\mathcal{C}_2 | Att(a_1)$ is often costly, because this set - even if smaller than $\mathcal{A}$ and $\mathcal{C}^t_2$ - may still be large. In order to take probabilities into account, the more efficient solution appears to \textit{generate randomly prototypes} from the matching BN, as was done for agent generation (see~\ref{indoc:generation_proto}), but this time with evidence on agent 1 attributes. The intuitive idea is to guess, given the knowledge on a mans' characteristics, what his girlfriend could look like. In the example of the spouses link, knowing that $a_1$ is a male agent 44-years-old living in R2, the generation will produce women that are on average 10 years younger and live in R2. Most generations will produce 34-years-old women, but some will propose less probable cases like 15, 17 or 43-years-old women. These prototypes are then searched in $\mathcal{C}_2 | Att(a_1)$. If a compatible agent $a_2$ is found, a link is created, and this link creation stopped. Else another prototype is generated, in order to explore the space of possible, and more or less probable agents $a_2$ in $\mathcal{C}_2 | Att(a_1)$.

The prototype-search approach works most of the time in an efficient way. It becomes less efficient when $\mathcal{C}_2 | Att(a_1)$ becomes small, because the probability to find an agent having well-defined characteristics becomes lower and lower when the set shrinks. This is especially true at the end of the generation process. For instance, when nearly all the men were linked with wives with optimal probabilities, it only remains some men that should be matched with few women. Exploring randomly the small space of $\mathcal{C}_2 | Att(a_1)$ by creating ideal candidates becomes a non-sense, that will test hundreds of probable combinations of attributes while only tens of less probable combination exist in the candidates set. When $\mathcal{C}_2 | Att(a_1)$ is too small, or that prototype-based search failed too much times, we shift to a \textit{fallback matching process}. The fallback process intuitively consists in iterating accross candidates and testing their compatibility. Agent $a_2$ is picked up randomly from $\mathcal{C}_2 | Att(a_1)$, its compatibility checked using the matching BN and a link created between $a_1$ and $a_2$. Node that this fallback solution can bias statistical distribution in the population; in our example, the fallback process may link several husbands with older wives, which is allowed from statistics but highly improbable. These errors will be statistically studied later (see \ref{indoc:errors}). When no fallback solution can be found, that is when $\mathcal{C}_2 | Att(a_1) = \{ \varnothing\}$, agent $a_1$ remains orphan, but will never be tied with a incompatible agent (in our example, no man will be said married with a non married or too young wife). 

% Evidence on attributes of $a_1$ and link creation $p(L^{t}_{a_1,a_2}=1)=1$ is still present in the matching BN, so the attributes of agent 2 are contrainst by $Att(a_1)$. 

\begin{figure}[t]
\begin{framed}
\begin{verbatim}
SELECT * FROM agents WHERE 1
AND agent_ageTranche='20-24'
AND lieu='village2'
AND (LK_friendship_requiredNb>=1 AND LK_friendship_requiredNb<=10)
AND (LKauto_friendship_createdNb>=0 AND LKauto_friendship_createdNb<10)
AND agent_id!=151
AND agent_id NOT IN (SELECT l1.agent_id1 FROM links l1 
                     WHERE l1.agent_id2=151)
AND agent_id NOT IN (SELECT l2.agent_id2 FROM links l2 
                     WHERE l2.agent_id1=151);
\end{verbatim}
\end{framed}
\caption{Example of SQL query automatically generated when searching for agents $a2$ candidate for linking as friends with a defined agent $a_1$ during the matching of spouses. Lines 2 to 5 are built from the state of evidence in the Bayesian Network, in order to select only agents compatible with the attributes of $a_1$. Line 6 excludes from results the agent $a_1$, that cannot be linked with itself. The four last lines exclude from the result all the agents that are already linked with $a_1$ with any type of link.}
\label{fig:example_SQL_select}
\end{figure}

The retrieving of subsets of the population given their attributes is facilitated by the storage of the population in a relational database. An example of automatically generated SQL query that enables to select the set $\mathcal{C}_2 | Att(a_1)$ from database is proposed in figure~\ref{fig:example_SQL_select}.

\subsection{Algorithm for creating transitive links}

\begin{figure}[ht]
\begin{framed}
\begin{verbatim}
SELECT l2.agent_id2 AS idFrom, l1.agent_id2 AS idTo 
FROM links l1, links l2 
WHERE l1.agent_id1 = l2.agent_id1
AND l1.link_type='motherOf' 
AND l2.link_type='married' 
AND l2.agent_id2 != l1.agent_id2
AND NOT EXISTS ( SELECT * FROM links l3     
                 WHERE l2.agent_id2 = l3.agent_id1 
                 AND l1.agent_id2 = l3.agent_id2)
AND NOT EXISTS ( SELECT * FROM links l3 
                 WHERE l1.agent_id2 = l3.agent_id1  
                 AND l2.agent_id2 = l3.agent_id2)
\end{verbatim}
\end{framed}
\caption{Example of automatically generated SQL query that retrieves the agents to link by transitivity, here for creating links ``fatherOf'' by transitivity of spouses and motherhood links. First line defines the fields that will be retrieved ($a_1$ and $a_2$). Line 2 and 3 defines the join between tables. Lines 4 and 5 constraint the kinds of links. The last lines exclude from results the dyads that were already linked by any link.}
\label{fig:sql_query_transitive}
\end{figure}

Transitive rules are defined as $p(L_{a_1,a_3}^{t_3}=1 | L_{a_1,a_2}^{t_1}=1, L_{a_2,a_3}^{t_2}=1) \in [0,1]$. Generating these transitive links is highly facilitated by the storage of the network into a database, because of the possibility to perform joins between tables, that is to deal efficiently with sets. We generate a SQL query that search database for all the agents $a_1$ and $a_2$ linked together with a link of type $t_1$, for which $a_2$ is also linked with an agent $a_3$ by a link $t_3$, with $a_1 \neq a_3$ and no link between $a_1$ and $a_3$ (an example of automatically generated query used for retrieving the agents to link by transitivity is shown in figure~\ref{fig:sql_query_transitive}). Once these results retrieved, we just add links of type $t_3$ between all dyads $a_1$ and $a_3$ with probability $p(L_{a_1,a_3}^{t_3}=1 | L_{a_1,a_2}^{t_1}=1, L_{a_2,a_3}^{t_2}=1)$. 

% Implementation adds minor variations to this principle to take into account directed and undirected links. 

\subsection{Implementation \& efficiency}

The core of the software is developed in Java. In order to improve the reliability and the efficiency of our generator, we preferred to rely on several external tools, each being specialized in a given task. The exact inference engine is based on the recursive conditioning algorithm \citep{samuel_thiriot:bib_apply_decision:darwiche_2001_1,samuel_thiriot:bib_apply_decision:allen_2003_1}.
Computations on Bayesian Networks is delegated to specialized library. \textit{Statistics on graphs} are computed using the statistical tool ``R'' \citep{samuel_thiriot:bib_software:Rteam_2009_1}, using the dedicated library iGraph \citep{samuel_thiriot:bib_software:csardi_2006_1}. Any \textit{relational database} compliant with SQL may be used for storage of the network and the population. We mainly use MySQL, a free and performant standalone database manager, and SQLite, which constitutes a zero-configuration database engine. The first enhance efficiency by processing requests on a dedicated process, enabling the usage of multi-core processors. The second enables an easier deployment of the software with acceptable performances. \textit{Networks visualization} is assumed by the graphviz software \citep{samuel_thiriot:bib_software:gansner_2000_1}.

The code was optimized in several places. For the \textit{SQL side}, indexes are automatically added in the SQL database for improving look-up for agents given their attributes. SQL queries were optimized for two main database managers, MySQL and SQLite, by changing the format of data (Enum types are more efficient in MySQL, but don't exist in SQLite) or by creating indexes on fields used in matching BN. These database managers propose various solutions to improve efficiency, including storage of tables in memory rather than in filesystem. \textit{Communication with the database engine} also slows the generation process. It is improved by caching insertion of agents, so only one query is used per tens of new agents, and by caching SQL queries for avoiding redundant operations. The \textit{inference engine} is implemented in the library named ``Ace''\footnote{Published in http://reasoning.cs.ucla.edu/ace/}, that encodes Bayesian networks into Conjunctive Normal Form for improving inference performances \citep{samuel_thiriot:bib_software:chavira_2006_1}. The choice of storing the network and the agents into a relational database also brings several benefits that include multitasking (the database engine takes in charge the look-up of the right information with optimal efficiency) and no limitation of the network size (no limit due to the memory available on the computer).

As expected, the generator works in a sublinear time compared to the number of possible links, even if more than linear compared to the size of the population. Agents generation ($\sim 0.2$ seconds per 100 agents) and creation of links transitivity ($\sim 0.01$ seconds per 100 agents) are both very quick and linear with $N$. The creation of links from matching is also very efficient ($\sim 0.20$ sec per 10 agents) at its beginning, when the candidate agents is large. However, we the sets of possible candidates becomes smaller, the computational time increases dramatically (up to $\sim 0.1$ second per agent) because of the number unsuccessful search for prototype agents (that also involve its generation) but also when searching agents with the fallback process. The overall generation process ranges from several minutes for small networks (less than 10000 agents) up to hours for large populations (100000 agents).

\subsection{Outputs}

The generator exports both the population and the network in various formats, in order to facilitate its use either for simulation or for analysis with specialized software. The multiplex network is exported as several uniplex networks (one per link type $t \in \mathcal{T}$) and as uniplex network (in a format that may be opened with UCINet, and graphviz. A graphical view of the multiplex network (if this one is not too large) is also exported in a vectorial format.

Other kinds of information are also exported. The Attribute BN learned on real data, that may be opened in a graphical editor, is exported in order to facilitate the perception of the bias in the generated population. Statistics on the graph (see~\ref{indoc:section_stats}) and on the generation process are also outputed.

\subsection{Quantifying errors\label{indoc:errors}}

\begin{figure}[t]\sidecaption
\includegraphics[width=0.6\textwidth]{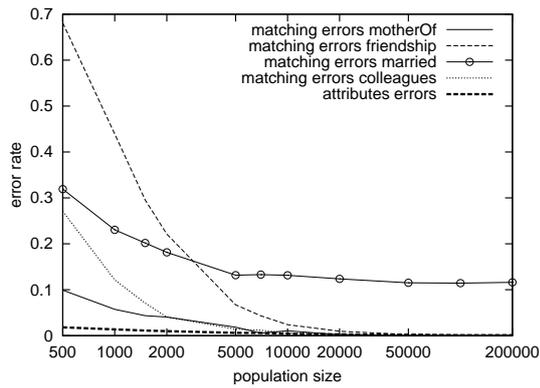} 
\caption{Error rate given population size.}
\label{fig:errors}
\end{figure} 

While BN describe a theoretical population using continuous probabilities, we generate a discrete population and link agents only when a suitable candidate exists. That limitation necessarily leads to a bias in the statistical properties of the population. Two kinds of errors may appear during generation. \textit{Errors on statistical distribution} appear because the generated population $\mathcal{A}$ is not large enough given the combinatorial of attributes' values described by the agent BN. These errors are measured by learning the agent BN on data, and quantified as the average difference between theoretical and measured probability. As shown in Fig.~\ref{fig:errors} \textit{(left)}, these errors (bottom curve) remain low and are negatively correlated to the population size. 

\textit{Errors on matching} appear when no candidate was found to link several agents or when all compatible agents were already linked, and are quantified as the rate of the total number of links required by $RC_t$ on the number of created ties. When that error rate remains low, and decreases when the population size increases, errors are only due to the discrete nature of agents: it will always exist some agents which could not be connected because their theoretical peer was not created. When the population is too small, it is also impossible to created as many links as desired: for instance, the algorithm cannot create $100$ links per agent in a population of less than $100$ agents. As shown in Fig.~\ref{fig:errors}, that error rate drops quickly above a given population size. Given our parameters, a population of 5,000 agents is a minimum to reduce errors. Above 10,000 agents, no more significant improvement appears. When the matching error rate remains high when the population size increases, it means that agent BN and/or matching BN are incompatible. 

\begin{framed}
Figure.~\ref{fig:errors} depicts the generation error measured for the Kenya case. Curve for link \textit{married} shows that the number of wives per men is not compatible with the proportion of married wives. In that case statistics (or assumptions) used to build BN should be checked and corrected. This error was shown on purpose, to underline the need to check these generation errors that may reveal an inconsistency in parameters. The error in link spouses is due to the fact that statistics for men where not consistent with those for women, that is there were more men that were expecting wives than women parametered to accept husbands.
\end{framed}

\section{Experiments and observations\label{indoc:experiments}}

\subsection{Generated Network}

\begin{figure}[t]
% \begin{tabular*}{\linewidth}{c c}
\includegraphics[width=12cm]{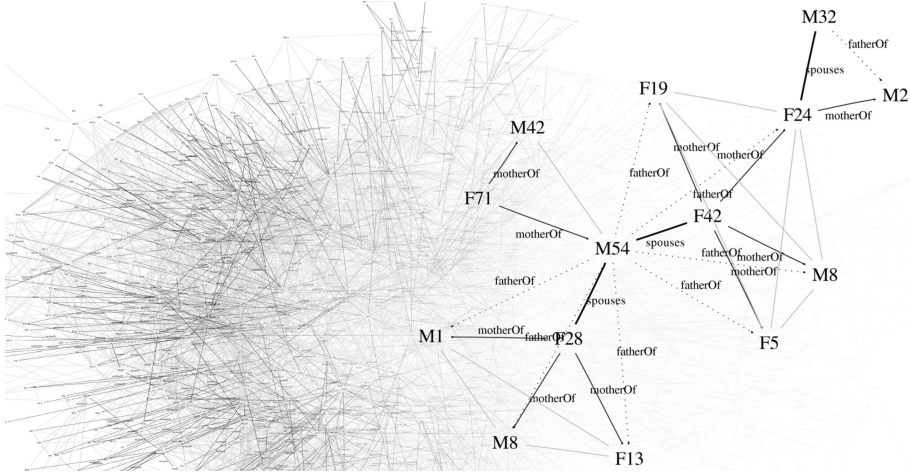} 
% &
% \includegraphics[width=6cm]{dot/householddef.jpg} \\
% \end{tabular*}
\caption{\textit{(left)} generated relationships network \textit{(right)} zoom in one agent}
\label{fig:result}
\end{figure}

The resulting graph includes of different types of links $t \in \mathcal{T}$, and provides the values of agent attributes $Att(a)$ for any agent $a \in \mathcal{A}$. Agents are placed on the network given their attributes. 
% using the generated network
The structure of relationships described by the generated network depends obviously on agent BN and matching BN provided by the modeler as parameters. In our application to Kenya, the population covers the whole age pyramid, and describes attributes depicted in Fig.~\ref{fig:attributesBN}. Moreover, as shown in Fig.~\ref{fig:result}, each agent is positioned in its familial environment; agent M54 (for Male, 54 years) is married with two wives F28 and F42, and has 7 children, including one daughter F24 who is herself married and mother. He is also tied with its own mother F71 and brother M42, but not with its father - probably because this one is not in the age pyramid (no more alive). He his also tied with \textit{colleagues} and \textit{friends} (not represented in that figure to improve lisibility). That structure is described at the scale of the 50,000 agents depicted in Fig~\ref{fig:result} \textit{(left)}.

To use that network for simulation, the modeler may simply define probability to interact given the kind of relationship: $\forall t \in \mathcal{T}$, $p^t_{interact}$, so  $p_{interact}(a_1,a_2 | L_{a_1,a_2}^t) = p^t_{interact}$. He may also choose a finer granularity by defining the probability of interaction given attributes $Att$, for instance to represent the fact that spatial distance decrease probability of interaction:  $p_{interact}(a_1,a_2 | L_{a_1,a_2}^t), Att(a_1),Att(a_2))$. Network use is part of the modelling task, that remains in charge of the modeler. We just help him/er by providing libraries that enable them to load the agents and the network for being used in a model.

% In that illustration, we focus on interactions about contraceptive use \cite{samuel_thiriot:bib_customer_value:rutenberg_1997_1}. In our case, no interaction occurs across links between young children and their parents. As the topic of contraceptive use is sensitive in Kenya, probabilities of discussion between spouses are low, as between a mother and its own parents. In fact, women which are still fertile and are concerned by the topic discuss mainly with their female friends, and often with their brothers-in-law (link sibling). The resulting network of interactions is a network in which ties are weighted by probabilities; it is by far sparse than the network of relationships.

\subsection{Properties of generated networks\label{indoc:section_stats}}

\begin{figure}[t]\sidecaption
\includegraphics[width=0.7\textwidth]{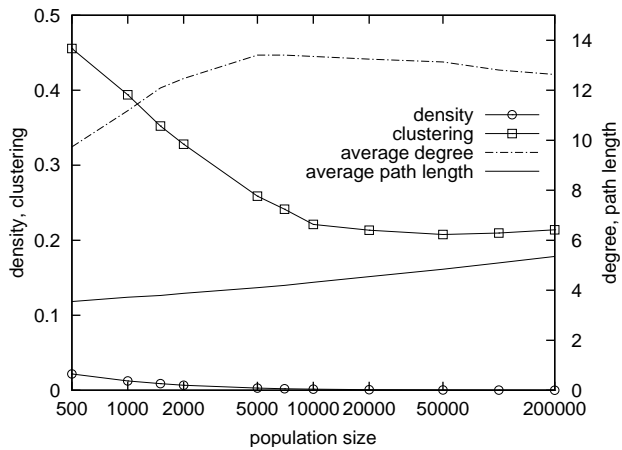}
\caption{Statistical properties of the generated relationship network in the case of Kenya.}
\label{fig:stats}
\end{figure}

\subsubsection{Statistical properties of networks}

Figure~\ref{fig:stats} \textit{(right)} depicts the evolution of statistical properties of the relationships network given the population size. We compute the classical network-scale statistics: \textit{density} is the number of links over the total possible links in an uniplex network of the same size, \textit{clustering} is the total number of triangles in the network \citep{samuel_thiriot:bib_sma_simulation:watts_1998_1}, \textit{average degree} is the total number of links divided by the population size, and the \textit{average path length} is the average geodesic distance between each pair of agents in the network. Note that these properties are computed on the uniplex network, all the types of social links taken together. Average path length is computed on the biggest cluster only. All of these values depend on the parameters. However, their behavior  is representative of the properties of graphs generated with our algorithm.

\textit{Density remains low} for any population size. It decreases when the population grows, because the number of links created for each agent is constant. 

The \textit{average degree} (dashed curve) directly depends on the parameters. The degree should theoretically be \textit{at least} the sum of the degree required for each link type t created by homophily. In fact, as explained before, all the required links cannot be created in a population smaller than 5000 agents. The bigger the population, the higher the number of links that can be created. This process explains why \textit{the degree curve only reaches its top for 5000 agents}. However, after 5000 agents, the curve begins a \textit{slow decrease when the population grows}. In fact, the degree depends on the two generative principles. The baseline degree is due to homophily: user defines the required number of links, and the algorithm creates this number of links. Once the population is large enough, this baseline degree becomes constant. However, transitivity adds more links depending to the number of already existing links. The number of links added to the baseline degree depends on transitivity, which itself decreases with the population size. 

\textit{High clustering emerges from homophily}: relationships often involve people sharing the same spatial location (colleagues, friends, spouses, parenthood) and similar age (friendship, spouses). Moreover, the creation of links by \textit{transitivity increases the clustering rate}: siblings obviously share the same parents, friends of friends become friends, etc. \textit{Common affiliations} also increase clustering, as people that share common workplaces have more chances to meet and bond. Note that transitivity starts from a very high value of $0.45$ (nearly each tryad of agents is closed), but stabilizes on a realistic value of $\sim 0.22$ when the population size exceeds 10000. Here is the principle that explains this second threshold: below 5000 agents, the population is so saturated that all the links cannot be created by transitivity. Between 5000 and 10000, the network is still very dense, so even if all the matching links can all be created (thus stabilizing measures on error rates and average degree), the network remains nearly saturated, as can be observed with the ``colleague'' error curve. Above 10000, this artificial density disappears, and nearly stabilizes for bigger population sizes.

The \textit{average path length} is both \textit{short} ($\sim 4,5$) and \textit{slowly increasing when the population grows}, characterizing the small-world phenomenon. The \textit{short average path length} appears because of the creation of some links that are not constrained by homophily, or by the creation of different kinds of social links that don't rely on the same attributes for homophily. In our illustration, family links between adults don't involve the same spatial area, as adults may change of town when they are old enough. In a similar way, friendship - if often occurring in the same spatial location - may cross large distances. All these social links create shortcuts between communities, exactly as explained by Watts and Strogatz in their small-world algorithm. 

Note that the distribution of degrees directly depends on the required degree per individual. If a preferential attachment is described into the parameters, then the distribution of degree in the network will exhibit a fat-tail distribution of degrees. 

% Additional links are also added by the transivity generation process. The curve first grows with the population size until the population reaches 5000. It is actually determined by two components. 

\paragraph*{}
Given analysis of both error and statistical properties of graphs given the population size, we shown that it exists a minimal population size required for generating a plausible population. This threshold depends on the complexity of the matching conditions passed as a parameter and of the number of links required in the matching BN passed as parameters.

\subsubsection{Emergent effects}

The purpose of the generator is to create a population that complies with parameters. We didn't expected any ``emergent'' effect during the generation. Some exist, however. For instance, many agents appeared to be linked to no family. The population was however compliant with available statistics, as were the statistics on the number of children per woman. It appeared that these agents were describing orphan children, who are numerous in Kenya due to AIDS. 
The use of Bayesian offers two benefits.

From the \textit{user viewpoint}, Bayesian Networks are graphical models that can be easily \textit{understood and manipulated}. As underlined by Judea Pearl, graphical models enable experts to focus on the qualitative model rather than the quantitative one \citep{samuel_thiriot:bib_ia_forte:pearl_1988_1}.  Bayesian networks, associated with software that simplify their visualization and manipulation, also constitute a way to check the coherency of parameters. For instance, many demographs viewing the attribute BN for Kenya asked us why no line links gender and age, while age pyramids are mostly asymmetrical (in fact, current pyramid in Kenya is a rare case of nearly-symmetrical age pyramid). Software like Samiam also enable users to ``play'' with Bayesian networks, by adding evidence in some values and observing the updated posterior probabilities. The lisibility and manipulability is a first importance in our approach, as it constitutes a way to retrieve information from field experts and to check the validity of parameters. 

From the \textit{developper viewpoint}, graphical models constitute a compact representation of a density of probabilities in a large space of variables. The activity of research on graphical models led to optimized algorithms and their implementation, which may be reused to optimize the efficiency of the generator. In our approach, Bayesian Networks were used with a retropropagation-capable inference engine that helped us to deal with large sets of agents in a reasonable time.

\section{Discussion}

\subsection{Summary}

Rather than sticking to existing practices for network generation, we analyzed this problematic with a modelling approach. We first defined the needs of users of network generators and their typical situation. We underlined that networks cannot be tackled as an uniform problematic, but should rather be viewed as a metaphor that is applied to very different natures of networks. We highlighted that modelers need to distribute agents in the network according to their attributes; they also would appreciate to distinguish the different kinds of social links in the network, in order to define more precisely the impact of interactions on agents. We pointed out the existence of data available as scattered statistics and qualitative observations which constitute the only data available on the structure of interactions. 

We propose another approach in which a multiplex network is build given rules observed from the field. In order to provide a generic algorithm usable as a tool for users, we proposed a methodology that helps users to formalize observations and a generative algorithm. The theory used in both the methodology and the algorithm is a simplification of the social selection processes undermined in sociology: generative rules are based on homophily (in its broader understanding) and transitivity. The methodology propose to encode individuals' attributes and matching rules with Bayesian Network, that are passed to the generator as parameters along with the transitivity rules. The generative algorithm creates randomly multiplex graphs, in which different types of relationships are described. Agents are positioned over the network given their attributes. Errors introduced during the generation of the graph are quantified, and the existence of a lower bound on the population size was highlighted. 

We illustrated the flexibility of this approach by generating the structure of interactions that impacts the diffusion of contraceptive solutions in rural Kenya. During this generation, we used no information but indirect observations from public statistics and published field studies. We generated a plausible family structure at the scale of thousands or tens of thousands agents. The multiplex network also describes friendship and colleagues social links. The analysis of networks generated underlined that they comply with the statistical observations on other social networks.

\subsection{Answers to typical questions}

\subsubsection{Is the generation process descriptive ?}

It is important to underline that \textit{our purpose is not to describe how networks appear}, as could appear as we use findings on how networks are created. Our only aim is to describe a network at time $t$ that is as descriptive as possible with few observations. The use of social selection processes to describe the probability of link existence is only choosen because we have some information on these processes. Our final purpose is limited to the description of the network, not of its generation.

\subsubsection{Why using Bayesian Networks ?}

The use of Bayesian offers two benefits.

From the \textit{user viewpoint}, Bayesian Networks are graphical models that can be easily \textit{understood and manipulated}. As underlined by Judea Pearl, graphical models enable experts to focus on the qualitative model rather than the quantitative one \citep{samuel_thiriot:bib_ia_forte:pearl_1988_1}.  Bayesian networks, associated with software that simplify their visualization and manipulation, also constitute a way to check the coherency of parameters. For instance, many demographs viewing the attribute BN for Kenya asked us why no line links gender and age, while age pyramids are mostly asymmetrical (in fact, current pyramid in Kenya is a rare case of nearly-symmetrical age pyramid). Software like Samiam also enable users to ``play'' with Bayesian networks, by adding evidence in some values and observing the updated posterior probabilities. The lisibility and manipulability is a first importance in our approach, as it constitutes a way to retrieve information from field experts and to check the validity of parameters. 

From the \textit{developper viewpoint}, graphical models constitute a compact representation of a density of probabilities in a large space of variables. The activity of research on graphical models led to optimized algorithms and their implementation, which may be reused to optimize the efficiency of the generator. In our approach, Bayesian Networks were used with a retropropagation-capable inference engine that helped us to deal with large sets of agents in a reasonable time. 

\subsubsection{A complex solution}

Modelers who daily use Barabasi-Albert or Watts-Strogatz generators first perceive this approach as much more complex. Explicative models can and have to be simple for reaching their purpose. On the contrary, descriptive models need more parameters and complexity for remaining connected with reality. That \textit{comparison seems us irrelevant}; in fact, this generator should rather be compared with other descriptive models of real networks, like p\textasteriskcentered models, that can hardly be applied in agent-based modelling. ``Simple, but not simpler'' \citep{samuel_thiriot:bib_sma_simulation:coen_2009_2}; the model as to be simpler than the reality, but shouldn't become too simple and failing to reach its purpose. Our model was built especially for helping modellers during their modelling process, and we tried to find a good tradeoff between complexity and descriptivity.

\subsubsection{While proposing a generic solution~?}

Proposing a generic solution may be perceived as useless by many computer scientists dealing with agent-based models. If they are interested in a specific model of interactions, they are obviously able to conceptualize and implement it by themselves. However, this argument ignores several of the needs of our research stream highlighted before (\ref{indoc:peculiar_needs_abm}). As explained before (\ref{indoc:approaches_abm}), an hand-made network generator can rarely be described in detail in the limited space of a paper. Such an ad-hoc solution thus limits the replicability of simulation results. Even if the generation process is detailed, its implementation may vary or contain errors; that wouldn't be the case of a generic algorithm. Our generator is parametrized by several files that may be shared along with the source code of simulation. Last but not least, new-comers in the field, or researchers having limited skills in computer science are not intersted/able to create their own generator. They require a tool, which is precisely what we described in this paper.

\subsection{Further work}

\subsubsection{Validation}

At this step, we already consider that generated networks are plausible. In fact, they are as descriptive as the data used as parameters. As these networks where build from observations, they already comply with available data. The graph-scale statistical properties were measured and shown to be realistic (as for the small-world phenomenon, for instance). Other statistical properties are tuned by parameters, thus cannot be used to validate the model. The only solution that would enable a better validation of this generator would be the comparison with a large dataset in which both agents' attributes and social links would be known. These datasets are quiet rare, but some (e.g. \citep{samuel_thiriot:bib_psycho:lewis_2008_1}) were recently proposed. We could try to rebuild the network given attributes, and check if some statistical indicators are different in the real and artificial network. The main problem would, however, to build the right statistical indicators to do so.

\subsubsection{Public diffusion}

The generator was built as a tool usable for any modeler. We plan to publish it as soon as possible on a free and open basis. Currently, the number of external softwares or libraries involved in the generator limit the communicability of the sourcecode for legal or technical reasons. We are working on rewriting parts of our software to limit dependencies to these libraries. 

Another point of importance is the optimization of the generative algorithm. We still investigate other solutions for speeding the generation process. Current generation delay is, however, already acceptable for common use. 

\subsubsection{Open research problematics}

The main purpose of our research was to propose a tool ready to use out-of-the box, in order to help modelers to create more descriptive networks. We are currently applying this approach to various cases, in order to detect limitations that may be induced by the homophily/transitivity generative principles. Applications to simulation in economics are in progress. We are currently investigating the \textit{simulation of various dynamics on generated networks} like information dynamics and diffusion of innovations.  Our approach enables the description of a realistic population at a large scale, thus enabling the reproduction of realistic marketing campains targetted on several segments. It could also be used to describe the structure of interactions in epidemics models, thus enabling to test the impact of various strategies to limit pandemics. 

Another possibility is to \textit{investigate the differences in generative rules that impact the collective dynamics}. These sensitivity studies should help us to collect more information on these rules for generating networks that would fully satisfy modelers. The \textit{study of the statistical properties of the network given the generation rules} is also a problematic of interest. 

As pointed out by several sociologists, \textit{generated networks constitute the first dataset of a very large network of relationships} that don't suffers from sampling biases. These networks could thus be used as a test for statistical indicators from sociology. For instance, it is the first time that the structure of the family is described at such a large scale. It is also the first time that we can generate various networks having so constrained by field observations. Generated networks could be used as tagged datasets, in which algorithms like communities detection could be test, by analyzing of the meaning and relevance of the communities they have detected given agents' attributes and generative rules.

\begin{acknowledgements}
Authors would like to thank Edmund Chattoe-Brown, Matthieu Latapay, Pierre-Henry Willemin, Camille Roth, Zach Lewckovitz for discussions on networks or on the algorithms presented in this paper. 
Part of this research was provided by research grant 2005/993 from ANRT (the French National Association for Technological Research) and by France Telecom R\&D - Orange Labs. Recent support was provided by RTRA STAE.
\end{acknowledgements}

\bibliography{../../commons/biblio_bibtex/bib_customer_value.bib,../../commons/biblio_bibtex/bib_customer_value_private.bib,../../commons/biblio_bibtex/bib_sma,../../commons/biblio_bibtex/bib_sma_organisation,../../commons/biblio_bibtex/bib_decision_psychology,../../commons/biblio_bibtex/bib_modelsimu,../../commons/biblio_bibtex/bib_model_decision,../../commons/biblio_bibtex/bib_psycho,../../commons/biblio_bibtex/bib_sma_simulation,../../commons/biblio_bibtex/bib_ethologie,../../commons/biblio_bibtex/bib_ia_forte,../../commons/biblio_bibtex/bib_general,../../commons/biblio_bibtex/bib_communications_privees,../../commons/biblio_bibtex/bib_perso,../../commons/biblio_bibtex/bib_apply_decision,../../commons/biblio_bibtex/bib_software}

\end{document}